\documentclass[preprint]{elsarticle}

\usepackage{graphicx}
\usepackage{float}
\usepackage{subcaption}
\usepackage{amsmath}
\usepackage{bm}
\usepackage{import}
\usepackage{color}
\usepackage{bm}
\usepackage[draft]{hyperref}
\usepackage{xcolor}

\definecolor{mycolor11}{RGB}{0,0,255}
\definecolor{mycolor12}{RGB}{15,0,240}
\definecolor{mycolor13}{RGB}{124,0,131}
\definecolor{mycolor14}{RGB}{255,0,0}

\definecolor{mycolor21}{RGB}{0,0,255}
\definecolor{mycolor22}{RGB}{2,0,253}
\definecolor{mycolor23}{RGB}{116,0,139}
\definecolor{mycolor24}{RGB}{255,0,0}

\definecolor{mycolor31}{RGB}{68,0,187}
\definecolor{mycolor32}{RGB}{0,0,255}
\definecolor{mycolor33}{RGB}{43,0,213}
\definecolor{mycolor34}{RGB}{255,0,0}

\definecolor{mycolor41}{RGB}{255,0,0}
\definecolor{mycolor42}{RGB}{195,0,60}
\definecolor{mycolor43}{RGB}{0,0,255}
\definecolor{mycolor44}{RGB}{60,0,195}

\definecolor{mycolor51}{RGB}{0,0,255}
\definecolor{mycolor52}{RGB}{6,0,249}
\definecolor{mycolor53}{RGB}{108,0,147}
\definecolor{mycolor54}{RGB}{255,0,0}

\definecolor{mycolor61}{RGB}{6,0,249}
\definecolor{mycolor62}{RGB}{8,0,255}
\definecolor{mycolor63}{RGB}{94,0,161}
\definecolor{mycolor64}{RGB}{255,0,0}

\definecolor{mycolor71}{RGB}{255,0,0}
\definecolor{mycolor72}{RGB}{115,0,140}
\definecolor{mycolor73}{RGB}{0,0,255}
\definecolor{mycolor74}{RGB}{115,0,140}

\definecolor{mycolor81}{RGB}{255,0,0}
\definecolor{mycolor82}{RGB}{195,0,60}
\definecolor{mycolor83}{RGB}{65,0,190}
\definecolor{mycolor84}{RGB}{0,0,255}

\definecolor{mycolor91}{RGB}{17,0,238}
\definecolor{mycolor92}{RGB}{0,0,255}
\definecolor{mycolor93}{RGB}{109,0,146}
\definecolor{mycolor94}{RGB}{255,0,0}

\definecolor{mycolor101}{RGB}{49,0,206}
\definecolor{mycolor102}{RGB}{0,0,255}
\definecolor{mycolor103}{RGB}{98,0,157}
\definecolor{mycolor104}{RGB}{255,0,0}

\definecolor{mycolor111}{RGB}{255,0,0}
\definecolor{mycolor112}{RGB}{148,0,107}
\definecolor{mycolor113}{RGB}{0,0,255}
\definecolor{mycolor114}{RGB}{255,0,0}

\definecolor{mycolor121}{RGB}{179,0,77}
\definecolor{mycolor122}{RGB}{255,0,0}
\definecolor{mycolor123}{RGB}{64,0,191}
\definecolor{mycolor124}{RGB}{0,0,255}

\begin{document}

\begin{frontmatter}

\title{A ``poor man's'' approach for high-resolution three-dimensional topology optimization of natural convection problems}

\author[mymainaddress]{Nicol\`o Pollini}
\author[mymainaddress]{Ole Sigmund}
\author[mymainaddress]{Casper Schousboe Andreasen}
\author[mymainaddress]{Joe Alexandersen\corref{mycorrespondingauthor}}
\ead{joealex@mek.dtu.dk}

\address[mymainaddress]{Section of Solid Mechanics, Department  of Mechanical Engineering, Technical University of Denmark, 2800 Kgs. Lyngby, Denmark}

\cortext[mycorrespondingauthor]{Corresponding author}

\begin{abstract}
This paper treats topology optimization of natural convection problems.
A simplified model is suggested to describe the flow of an incompressible fluid in steady state conditions, similar to Darcy's law for fluid flow in porous media. The equations for the fluid flow are coupled to the thermal convection-diffusion equation through the Boussinesq approximation.
The coupled non-linear system of equations is discretized with stabilized finite elements and solved in a parallel framework that allows for the optimization of high resolution three-dimensional problems.
A density-based topology optimization approach is used, where a two-material interpolation scheme is applied to both the permeability and conductivity of the distributed material.
Due to the simplified model, the proposed methodology allows for a significant reduction of the computational effort required in the optimization. At the same time, it is significantly more accurate than even simpler models that rely on convection boundary conditions based on Newton's law of cooling.
The methodology discussed herein is applied to the optimization-based design of three-dimensional heat sinks. 
The final designs are formally compared with results of previous work obtained from solving the full set of Navier-Stokes equations.
The results are compared in terms of performance of the optimized designs and computational cost.
The computational time is shown to be decreased to around $5-20\%$ in terms of core-hours, allowing for the possibility of generating an optimized design during the workday on a small computational cluster and overnight on a high-end desktop.
\end{abstract}

\begin{keyword}
topology optimization \sep conjugate heat transfer \sep natural convection \sep high performance computing \sep simplified model
\end{keyword}

\end{frontmatter}


\section{Introduction}
\label{sec:intro}
Convection is the heat transfer due to movement of a fluid.
Two types of convection can be identified: natural convection, when the fluid flow is caused by buoyancy forces generated by temperature gradients in the fluid; and forced convection, when the fluid flow is caused by an external forcing.
Both forced and natural convection are considered in the design of heat sinks. 
These are heat dissipation devices often used to enhance the rate of dispersion of heat generated by different power sources within computers, other electronic enclosures or heat producing machinery. 
Heat sinks typically consist of a highly conductive material surrounded by a fluid. 
In the design of heat sinks, the dimensions may be constrained by product size requirements, and in many cases the fin array geometry is one of the few parameters left at the engineer's discretion.
In order to obtain innovative designs of heat sinks with improved heat transfer, several researchers focused on the optimization-based design of heat sinks.
For example, \citet{Morrison} focused on the optimization of the fin geometry of heat sinks in natural convection with rectangular cross-section fins at a constant spacing. A derivative-free approach was used, where the design variables were the fin and backplate thicknesses, and the fins' spacing. \citet{Ledezma1997} optimized the geometry of staggered vertical plates with the objective of maximizing the thermal conductance between the plates' assembly and the surrounding fluid. The least-material optimization of pin-fin, plate-fin and rectangular fin heat sinks in natural convection was discussed by \citet{Iyengar2003}. \citet{Bahadur2005} discussed the optimization-based design of vertical pin-fin heat sinks for a microprocessor cooled by natural convection. The design involved the pin height, diameter, and spacing. \citet{Jang2012} used a genetic algorithm to optimize a pin-fin radial heat sink.
The above mentioned references are a few examples of optimization approaches for natural convection systems based on parametric models with few design variables.
To avoid the need of predefining any aspect of the design problem beforehand, thus allowing for a large design space, this work considers a topology optimization approach.

Topology optimization can be considered the most general form of structural optimization  that allows for vast design freedom \cite{Bendsøe2003}. 
The basic idea consists of distributing a given amount of material within a prescribed design domain in order to obtain the best structural performance. 
It often results in highly efficient but unpredictable optimized designs, that could not have been obtained by simple intuition. Since the seminal paper by \citet{bendsoe1988generating}, topology optimization has undergone a tremendous development in several directions and fields \citep{Sigmund2013,Deaton2014b}. 
For example, topology optimization has been applied to conjugate heat transfer problems with forced convection  \cite{Yoon2010,Matsumori2013,Marck2013,Koga2013,Makhija2015a,
Yaji2015,Haertel2018a}. 
Recently, topology optimization has been applied also in the context of latent heat storage system design with phase change material \citep{Pizzolato2017}, complex turbulent flow systems \citep{Dilgen2018a}, and turbulent forced convection \citep{Dilgen2018}.

To this date topology optimization for fully-coupled natural convection problems has received only little attention.
Topology optimization for 2-D natural convection problems was first applied by \citet{Alexandersen20142D}, considering a density-based approach for the design of heat sinks and micro-pumps. The problem is formulated under the assumption of steady-state laminar flow, using the incompressible Navier-Stokes equations coupled to the convection-diffusion equation through the Boussinesq approximation. 
\citet{Alexandersen2016} later extended the formulation to the case of large scale 3-D topology optimization of heat sinks. This approach has been applied to the design of passive coolers for light-emitting diode (LED) lamps \cite{Alexandersen2018}. It is shown that topology optimization successfully identifies innovative design solutions that prove to outperform those achieved by intuition. 
In the context of topology optimization based on the level-set method \citep{VanDijk2013}, \citet{Coffin2016a} consider both steady-state and transient natural convection problems.
The experimental validation of topology optimized devices for conjugate heat transfer problems has been rarely performed, but there are a few contributions \cite{Dede2013a,Koga2013b,LazarovBS2018,Lei2018}.
In particular, \citet{LazarovBS2018} presents the experimental validation of the numerical results from \citet{Alexandersen2018}  using additive manufacturing in aluminium, closing for the first time the design-validation-manufacturing cycle for topology optimization of heat sinks passively cooled by natural convection. Through numerical and experimental results, it is shown that topology-optimized and additively-manufactured designs for passive coolers of LED lamps lead to significant material savings and performance improvements with respect to lattice designs.
Further experimental validation is discussed by \citet{Lei2018}. In this case, stereolithography-assisted investment casting (SLA-assisted IC) is used to fabricate heat sink devices designed through topology optimization. It is shown that SLA-assisted IC is a valuable alternative to more traditional metal additive manufacturing, and that it requires lower costs and is more flexible with regards to part size and metals that can be used.
However, even though the above mentioned approaches for topology optimization of natural convection problems accurately capture the physical description of the fluid flow through the Navier-Stokes equations, they also require a very high computational effort and time. In fact, a current limitation for a broader adoption of topology optimization in the early stage design of natural convection problems is the high computational effort that it requires.
For example, \citet{Alexandersen2016} report a computational time of approximately $10$ hours using $1280$ CPUs for topology optimization of natural convection problems. 

In an attempt to reduce the computational cost required for topology optimization of natural convection problems, several authors adopted a simplified convection model for the heat flux at the solid-fluid interface based on Newton's Law of Cooling (NLC). This simplified approach does not require the solution of the flow field, significantly reducing the computational cost required. 
For example, the NLC model has been applied in 2-D density-based topology optimization problems \cite{Sigmund2001b,Yin2002,Bruns2007,Alexandersen2011,Zhou2016}.
\citet{Coffin2016} consider the NLC model for topology optimization of 2-D and 3-D convective heat transfer problems through an explicit level-set method. 
They also observe that even though the NLC model approximates convective fluxes at the solid-fluid interface with a simple and computationally efficient formulation, it may significantly over-predict the heat flux, thus promoting the formation of thin fluid channels in the optimized topologies. They also report that this behavior can be mitigated, if the temperature field in the fluid is approximated by a diffusion model, and that the formation of small  solid elements is prevented with an explicit feature size control.
\citet{Joo2017a} extends the simple NLC model to account for the shape-dependency of the heat transfer coefficient in the problem formulation. The results show that the use of a shape-dependent variable definition of the heat transfer coefficient accounts with more accuracy for the actual channel spacing between the fins. This approach has recently been extended to the 3-D case \cite{Joo2018}, where the heat transfer coefficient depends on the local shape of the fins and decreases along the direction of the flow. 

An alternative to the NLC method for reducing the computational cost in natural convection problems is to consider a simplified flow model resembling Darcy's law for fluid in porous media.
Darcy's law is used by \citet{Guest2006a} for topology optimization of creeping fluid flows.
In particular, the Darcy equation is used to model the flow in the solid domain,  thus allowing to formulate the flow with a unified Darcy-Stokes flow formulation on a given solid-fluid domain.
A simplified flow model based on Darcy's law is also considered by \citet{Zhao2018} for 2-D topology optimization of turbulent forced convection for cooling channels.
Similarly, \citet{Asmussen2019} proposed a methodology for the topology optimization of 2-D heat sinks cooled by natural convection. A potential flow model, resembling Darcy's law, is considered to simplify the fluid flow, leading to promising optimized designs that are in good agreement with those achieved considering the full Navier-Stokes flow model. It is also shown, that the approach yields significantly better designs than an approach based on NLC. Both in the work by \citet{Zhao2018} and \citet{Asmussen2019}, particular attention is given to the definition of the numerical values of the fictitious fluid permeability parameters required by Darcy's law. In particular, \citet{Zhao2018} compares the simplified flow model to a full-blown turbulent flow benchmark example. The average velocity at a predefined cross-section of the domain, the temperature distribution at the center of the heat source and the pressure drop from inlet to outlet are used to perform the parametric tuning. 
\citet{Asmussen2019} perform a tuning procedure by comparing the temperature distribution associated with the simplified flow model and the incompressible Navier-Stokes flow in a benchmark example.

\begin{figure}
    \centering
\includegraphics[width=\columnwidth]{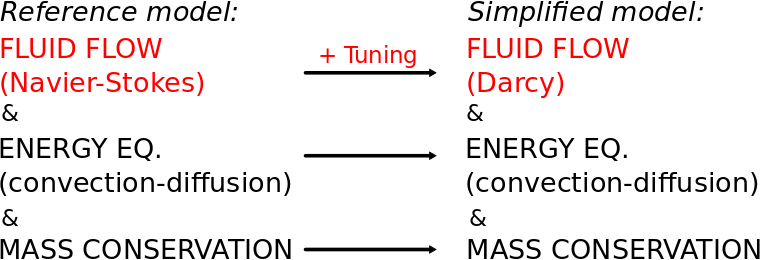}
    \caption{Visual description of the simplified natural convection model considered.}\label{fig:simplemodelimage}
\end{figure}

In this work, we extend the approach presented by \citet{Asmussen2019} to topology optimization of 3-D high resolution natural convection problems. The simplified flow model resembling Darcy's law is coupled to the thermal convection-diffusion equation through the Boussinesq approximation (Fig. \ref{fig:simplemodelimage}). We propose a tuning procedure for the fictitious fluid permeability parameter following analytical results available in the literature. In the solid region, we mimic the absence of flow by assigning an infinitesimal value to the permeability of the solid material. The non-linear system of governing equations is solved using stabilized finite elements in a parallel high-performance computing framework that allows for optimizing large scale problems. A density-based topology optimization approach is used, where a two-material (i.e. solid-fluid) interpolation scheme is applied to both the permeability and conductivity of the distributed material. 
As a consequence of the simplified flow formulation considered, the proposed approach allows for a significant reduction of the computational effort required in the optimization, and at the same time it is significantly more accurate than even simpler models based on NLC, as already shown by \citet{Asmussen2019}. The methodology discussed herein is applied to the optimization of academic heat sink design cases. The results are  compared to those presented by \citet{Alexandersen2016}, obtained considering an equivalent problem formulation using the full Navier-Stokes flow model. The comparison is done in terms of computational cost, and performance of the optimized designs.
Moreover, we suggest the use of the simplified natural convection model discussed herein in conjunction with a more accurate one based on the Navier-Stokes flow model. The result is a hybrid optimization approach, that leads to optimized designs with a superior performance compared to those obtained considering the simplified model only, but requiring a reasonable computational cost.  

The remainder of the article is organized as follows: Sec. \ref{sec:probform} presents the formulation of the problem at hand, with details on the governing equations of the problem; Sec. \ref{sec:tune darcy} provides a detailed description of the proposed tuning procedure for the fluid permeability parameter; Sec. \ref{sec:weakform} briefly discusses the finite element approximation adopted for the evaluation of the system response; Sec. \ref{sec:topopt} formulates the topology optimization problem considered with details on the material interpolation technique adopted and on other computational aspects; Sec. \ref{sec:numapp} presents an in-depth study and comparison for the benchmark example of \cite{Alexandersen2016}; Sec. \ref{sec:secondExample} presents a second numerical example illustrating the strengths and limitations of the proposed approaches; lastly, concluding remarks are given in Sec. \ref{sec:end}.

\section{Problem formulation}
\label{sec:probform}
In this section we present the governing equations. 
In particular, we first present the procedure adopted to simplify the fluid flow model.
This has recently been presented by \citet{Asmussen2019}, but is reproduced here for completeness.
We then conclude this section presenting the full set of governing equations including the incompressibility condition, and the equations for the energy conservation.

\subsection{From Navier-Stokes to Darcy} 
\label{sec:derivation}

We begin by considering the Navier-Stokes equations for a laminar and incompressible steady-state fluid flow. Through a sequence of assumptions, the final simplified flow model is obtained, and it can be defined also in terms of Darcy's law.
The steady-state incompressible Navier-Stokes equations considered are the following:
\begin{equation}
\rho u_{j}\frac{\partial u_i}{\partial x_j} - \mu \frac{\partial}{\partial x_j}\left( \frac{\partial u_i}{\partial x_j} + \frac{\partial u_j}{\partial x_i} \right) + \frac{\partial p}{\partial x_i} - \rho g_{i} = 0 \label{eq:app_NS}
\end{equation}
where $u_{i}$ is the $i$-th component of the velocity vector $\textbf{u}$, $\rho$ is the density, $\mu$ is the fluid viscosity, $p$ is the pressure, and $g_{i}$ is the $i$-th acceleration component of the gravity vector $\textbf{g}$.

The equations that model the fluid flow are coupled with the convection-diffusion equations through the Boussinesq approximation. The Boussinesq approximation links the fluid density with its temperature assuming a linear formulation, and it mimics the occurrence of buoyancy due to differences in the density caused by differences in temperature:
\begin{equation}
\rho = \rho_{0} - \alpha \rho_{0}(T-T_{0}) = \rho_{0} - \alpha \rho_{0}\Delta T
\end{equation}
where $\alpha$ is the coefficient of thermal expansion, $\rho_{0}$ and $T_{0}$ are reference values of the density and temperature. 
For convenience, we define the variable $P$, which is the modified pressure measure which includes the gravitational head:
\begin{equation}
P = p - \rho_{0}g_{i}x_{i}
\end{equation}

If we assume that the buoyancy term is dominant compared to the inertia term,  or more explicitly:
\begin{equation}
\left| \rho_{0} u_{j}\frac{\partial u_i}{\partial x_j} \right| \ll \left| \rho_{0}\alpha  \Delta T g_{i} \right| \label{eq:app_negligble}
\end{equation}
then Eq. \eqref{eq:app_NS} becomes:
\begin{equation}
- \mu \frac{\partial}{\partial x_j}\left( \frac{\partial u_i}{\partial x_j} + \frac{\partial u_j}{\partial x_i} \right) + \frac{\partial P}{\partial x_i} + \rho_{0}\alpha \Delta T g_{i} = 0  \label{eq:app_midpoint}
\end{equation}
Last, we consider the viscous resistance term to be linearly dependent on the velocity:
\begin{equation}
\mu \frac{\partial}{\partial x_j}\left( \frac{\partial u_i}{\partial x_j} + \frac{\partial u_j}{\partial x_i} \right) \approx -\frac{\mu}{\kappa} u_i
\label{eq:visclinear}
\end{equation}
where $\kappa$ is the permeability parameter associated to a fictitious porous medium. By substituting Eq. \eqref{eq:visclinear} into Eq. \eqref{eq:app_midpoint} we obtain the simplified flow model, that resembles Darcy's law for fluid flow in porous media:
\begin{equation}\label{eq:finaldarcy}
u_i = - \frac{\kappa}{\mu}\left(  \frac{\partial P}{\partial x_i} + \rho_{0}\alpha \Delta T g_{i}\right)
\end{equation}

\subsection{Governing equations} \label{sec:goveqs}
\begin{figure}
\centering
  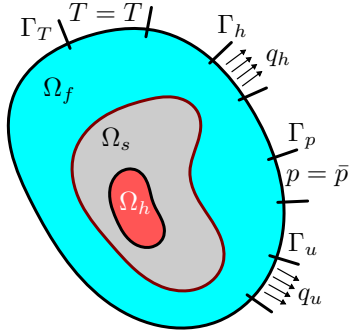
\caption{Generic representation of the systems considered. $\Omega_{s}$ is the design domain, $\Omega_{f}$ is the fluid domain, $\Omega_{h}$ is a solid domain portion with volumetric heat source and it is not involved in the design.}
\label{fig:gausslobatto1}      
\end{figure}
The governing equations are written for a unified domain, that includes both the solid and fluid part, i.e. $\Omega = \Omega_{s} \cup \Omega_{f}$. The characterization of the governing equations to each of the two sub domains is achieved by controlling the parameters that define the material behavior. In particular, we will interpolate the material permeability, $\kappa$, and conductivity, $k$, between the two domains (i.e. solid and fluid).
The following are the governing equations for the simplified flow model, incompressibility condition, and energy conservation:
\begin{equation}
\begin{split}
& u_{i} = - \frac{\kappa}{\mu} \bigg( \frac{\partial P}{\partial x_{i}} + \rho_{0}\alpha g_{i}\Delta T \bigg)\\
& \frac{\partial u_{i}}{\partial x_{i}} = 0 \\
& \rho_{0} c_{p} u_{i} \frac{\partial T}{\partial x_{i}} -\frac{\partial}{\partial x_{i}}\left( k \frac{\partial T}{\partial x_{i}} \right) - Q= 0
\end{split}
\label{eq:goveq}
\end{equation}
where $c_{p}$ is the specific heat capacity, $T$ is the temperature, and $Q$ is the volumetric heat source. Further, the following boundary conditions hold:
\begin{equation}
\begin{split}
 \text{Thermal:} \quad & \quad T=\bar{T} \quad \text{on} \quad \Gamma_{T}\\
& \quad k\frac{\partial T}{\partial x_{i}}n_{i} = q_{h} \quad \text{on} \quad \Gamma_{h}\\
 \text{Fluid:} 
\quad & \quad p=\bar{p} \quad \text{on} \quad \Gamma_{p}\\
& \quad u_{i}n_{i}=q_{u} \quad \text{on} \quad \Gamma_{u}
\end{split}
\end{equation}
When solving the above equations, the explicit expression for the velocity components are inserted into the incompressibility condition to give a Poisson-type equation for the pressure. Thus, the number of degrees-of-freedom (DOFs) is reduced to only 2 for the simplified model, namely the pressure, P, and the temperature, T. This is in contrast to the full Navier-Stokes model, where the number of DOFs are 5, namely the three velocity components, $u_i$, in addition to the two aforementioned. This is what yields the main computational reduction of the proposed approach.

Both $\kappa$ and $k$ are theoretically varying between solid (i.e. $\kappa_{s}$ and $k_{s}$) and fluid (i.e. $\kappa_{f}$ and $k_{f}$) properties:
\begin{equation}\label{eq:interpkK}
\left\{\kappa(\textbf{x}),\,k(\textbf{x})\right\} =\left\{
\begin{aligned}
& \left\{\kappa_{s},\, k_{s}\right\}\quad \text{if} \;\; \textbf{x}\in \Omega_{s}\\
&\left\{\kappa_{f},\, k_{f}\right\} \quad \text{if} \;\; \textbf{x}\in \Omega_{f}
\end{aligned}\right.
\end{equation}
Because the convection term of the energy equation in \eqref{eq:goveq} vanishes in the solid domain, $c_{p}$ has a constant value associated to the fluid phase.
The heat source $Q$ is assigned only to the solid domain $\Omega_{h}$, which is not involved in the design. Thus, also $Q$ will not be considered design dependent.
The interpolation of the two parameters between the two materials in Eq. \eqref{eq:interpkK} is defined through the design field $\gamma(\textbf{x})\in [0,1]$: 
$\gamma(\textbf{x})=0$ indicates that the element with coordinates $\textbf{x}$ belongs to the fluid domain; $\gamma (\textbf{x})=1$ means that the element with coordinates $\textbf{x}$ belongs to the solid domain. The response of the systems considered herein will be evaluated numerically with a finite element approach, where to each element $i$, a design variable $\gamma_{i}$ will be assigned. The vector $\bm{\gamma}$ collects all the design variables. 
Only the variables $\gamma_{i}$ associated with elements belonging to the subdomain $\Omega_{s}$, with the exclusion of $\Omega_{h}$, will be actual design variables of the problem.

\section{Tuning of the artificial fluid permeability}
\label{sec:tune darcy}

Darcy's law was originally conceived to describe
 the flow of a fluid through a porous medium, characterized by a permeability parameter.
In this work, we use it to emulate true fluid flow. 
For this reason, the permeability parameter $\kappa$ is artificial and needs to be tuned in order to adequately model the flow of the pure fluid considered, as well as the absence of flow in the solid domain.
 The topology optimization problems considered in this work will be characterized by a two-material distribution, i.e. a solid material and a fluid material.
 Therefore, the permeability parameters of these two materials (i.e. $\kappa_{s}$ and $\kappa_{f}$) need to be appropriately defined.
 
\begin{figure}\centering
	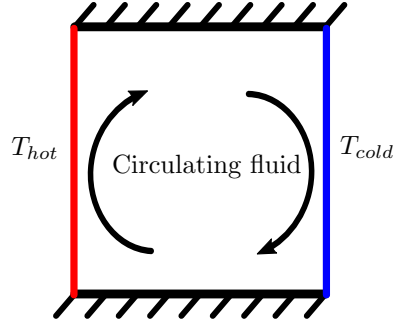
	\caption{Representation of the 2-D enclosure considered in the tuning procedure for the fluid permeability $\kappa_{f}$ described in Sec. \ref{sec:tune darcy}, with $\Delta T = T_{hot}-T_{cold}$.}
\label{fig:box2Dtune} 
\end{figure}

Ideally, the permeability of the solid domain is zero, but will in what follows be set to a very small number to avoid numerical issues (e.g. $\kappa_{s}=10^{-7}\, m^{2}$).
 The permeability of the fluid $\kappa_{f}$ is tuned considering analytical results available in the literature for a 2-D fluid cell recirculating inside a rectangular enclosure \citep{Bergman2011,Nield2017}. 
In particular, we consider the case in which the enclosure contains pure fluid (i.e. governed by the Navier-Stokes equations) and fluid in a porous medium (i.e. governed by Darcy's law). The vertical walls are isothermally heated, while the horizontal ones are thermally insulated (Fig. \ref{fig:box2Dtune}). 
Between the two vertical plates, the same fixed $\Delta T$ has been considered in the two cases. The tuning procedure consists in assuming that in both cases there is equal heat flux exchanged between the two vertical plates of the enclosure, translated to equal average Nusselt numbers. 
The case of pure fluid recirculating in a rectangular enclosure is discussed by \citet{Bergman2011}. In this case, the Nusselt number can be calculated as follows:
\begin{equation}\label{eq:nusselt1}
\begin{split}
&{Nu} = 0.18 \left( \frac{Pr}{0.2+Pr}Ra_{H}\right)^{0.29}
\end{split}
\end{equation}
where
\begin{equation}\label{eq:nusselt2}
\begin{split}
& Ra_{H} = Gr_{H}\,Pr, \;
Gr_{H}= \frac{g\, \alpha\, \Delta T \, H^{3}}{\nu ^{2}}, \;
Pr = \frac{\nu}{\beta}
\end{split}
\end{equation}
In Eq. \eqref{eq:nusselt1} and Eq. \eqref{eq:nusselt2}, ${Nu}$ is the average Nusselt number of the vertical plate with height $H$; $Pr$ is the Prandtl number; $Ra_{H}$ is the Rayleigh number of the vertical plate with height $H$; $Gr_{H}$ is the Grashof number; $g$ the gravity acceleration constant; $\alpha$ the fluid thermal expansion coefficient; $H$ the height of the enclosure; $\nu$ the fluid kinematic viscosity; and $\beta$ the fluid thermal diffusivity.

Similarly, the case of clockwise side-to-side convection in an enclosure with a porous medium is discussed by \citet{Nield2017}.
In this case, the Nusselt number can be approximated as follows:
\begin{equation}\label{eq:nusselt3}
\begin{split}
&Nu \approx \frac{L}{H} Ra^{0.5}
\end{split}
\end{equation}
with
\begin{equation}\label{eq:nusselt4}
\begin{split}
&Ra =\frac{g\, \alpha\, \kappa \, H\,\Delta T }{\nu \, \beta_{m}}
\end{split}
\end{equation}
where $\kappa$ is the permeability; $L$ the distance between the vertical plates; and $\beta_{m}$ is the thermal  diffusivity of the porous medium.
If we equate Eq. \eqref{eq:nusselt1} and Eq. \eqref{eq:nusselt3}, and then rewrite the resulting expression explicitly for $\kappa$ we obtain the following relation:
\begin{equation}\label{eq:nusselt5}
\begin{split}
&\kappa_{f} =  \frac{\beta_{m}\,H^{1.74}}{\nu^{0.16}\,(g\, \alpha\, \Delta T)^{0.42}\,L^{2}}\,0.0324 \left( \frac{Pr^{2}}{0.2+Pr}\right)^{0.58} 
\end{split}
\end{equation} 
where the subscript $f$ indicates that the estimated permeability refers to the fluid domain.

In Sec. \ref{sec:numapp}, we will consider four design cases discussed by \citet{Alexandersen2016}\footnote{Herein all quantities are given with units. However, the values are set so as to be equivalent to the non-dimensional quantities presented by \citet{Alexandersen2016}.}. They are characterized by a cubic domain with dimensions $1.0\,m\times1.0\,m\times1.0\,m$. Due to the symmetry of the problem we expect that for a generic design there will be rolling convective cells of circulating fluid occupying a quarter of the domain.
Thus, in Eq. \eqref{eq:nusselt5} we assume $H=1.0\,m$ and $L=0.5\,m$.
 Moreover, we will tune the fluid permeability specifically for each design case. Hence, the following values of the coefficient of thermal expansion will be considered: $\alpha=\{10^{3},10^{4},10^{5},10^{6}\}\,K^{-1}$. The values of $\Delta T$ adopted in Eq. \eqref{eq:nusselt5} will be based on the results discussed by \citet{Alexandersen2016}. However, similar results can be obtained considering $\Delta T=1\,K$. All the remaining parameters are set to one.
 
 The results of the tuning procedure are listed in Table \ref{tab:2}.
\begin{table}\centering\small
\caption{Values of $\kappa_{f}$ obtained with the tuning procedure presented, i.e. Eq. \eqref{eq:nusselt5}.}
\label{tab:2}       
\begin{tabular}{lllll}
\hline\noalign{\smallskip}
$\alpha$ $[K^{-1}]$ & $10^{3}$ & $10^{4}$ &  $10^{5}$ & $10^{6}$ \\
\hline\noalign{\smallskip}
$\Delta T$ $[K]$  & $1.7$ & $1.5$ & $1.2$ & $0.9$\\
\hline\noalign{\smallskip}
$\kappa_{f}$ $[m^{2}]$  & $0.00513$ & $0.00206$ & $0.00085$ & $0.00036$ \\
\hline\noalign{\smallskip}
\end{tabular}
\end{table}
It can be observed that the estimated fluid permeability gets smaller for increasing values of the fluid thermal expansion coefficient.
In the numerical applications in Sec. \ref{sec:topopt} we will consider the fluid permeability parameters listed in Table \ref{tab:2}.

\section{Finite element formulation}
\label{sec:weakform}
The governing equations are discretized using trilinear hexahedral finite elements. To obtain a smooth and non-oscillatory numerical solution of the governing equations, stabilization terms are added to the weak form of the convection-diffusion equation. 
The details of the formulation can be found in \ref{app:FEM}.

\section{Topology optimization}
\label{sec:topopt}

In this section we first present the main topology optimization problem formulation. Subsequently, we provide additional information regarding the adopted material interpolation functions, continuation scheme, and other computational considerations. 

\subsection{Optimization problem}
\label{sec:optprob}
The goal of the optimization is to find an optimized material distribution that minimizes an objective function with a constraint on the maximum amount of solid conductive material.
The objective function minimized is the thermal compliance:
\begin{equation}\label{eq:objfunc}
f(T) = \int_{\Omega_{h}} Q\, T d\Omega
\end{equation}
Considering the numerical approximation introduced for evaluating the response of the system (Eq. \eqref{eq:govequation} of \ref{app:FEM}), the discretized version of the objective function actually minimized is the following:
\begin{equation}\label{eq:objdiscr}
f(\textbf{s})= \sum_{e=1}^{N_{ele}}\textbf{f}_{t\,Q,\,e}^{T}\,\textbf{t}_{e}
\end{equation}
where $N_{ele}$ is the number of finite elements.
We impose a constraint on the volume of the solid domain:
\begin{equation}\label{eq:constr}
g(\bm{\gamma})= \frac{\sum_{e=1}^{N_{\mathcal{D}}}\gamma_{e}v_{e}}{\sum_{e=1}^{N_{\mathcal{D}}}v_{e}} - g^{*} \leq 0
\end{equation}
where $N_{\mathcal{D}}$ is the number of elements included in the design domain $\mathcal{D}$ (with $\mathcal{D}=\Omega_{s}\backslash \Omega_{h}$), $v_{e}$ is the volume of element $e$, and $g^{*}$ is the volume fraction of the solid domain with respect to the full design volume of the system considered.
The optimization problem can then be written as:
\begin{equation}\label{eq:optprob}
\begin{split}
 \underset{\gamma\in \mathcal{D}}{\text{minimize:}} &\quad f\left(\textbf{s}(\bm{\gamma}),\bm{\gamma}\right) \\
\text{subject to:} & \quad g(\bm{\gamma})\leq 0\\
\text{with:}& \quad \textbf{R}(\textbf{s}(\bm{\gamma}),\bm{\gamma})=0\\
& \quad 0 \leq \gamma_{i} \leq 1, \; i=1,\ldots,N_{ele}
\end{split}
\end{equation}
The topology optimization problem \eqref{eq:optprob} has been solved with an iterative gradient-based optimization algorithm, namely the  Method of Moving Asymptotes (MMA) \citep{Svanberg1987,Aage2013,Aage2014}. This algorithm relies on first order information. Therefore, the gradients of the objective function and of the constraint need to be calculated. The constraint function (i.e. $g(\bm{\gamma})$) is formulated explicitly in terms of the design variables. Hence, its gradient can be calculated explicitly. On the contrary, the dependency of the objective function (i.e. $f\left(\textbf{s}(\bm{\gamma}),\bm{\gamma}\right)$) on the design variables (i.e. $\bm{\gamma}$) is expressed implicitly through a non-linear relation. For this reason, we rely on adjoint sensitivity analysis to calculate the objective function gradient.

In the optimization analysis, the design field is regularized through a PDE-based (partial differential equation) filter to avoid the appearance of checkerboard patterns \citep{Lazarov2011,Aage2014}.

\subsection{Material interpolation}
\label{sec:RAMP}
The purpose of the optimization-based design approach discussed herein is to identify an optimized topology described by $\bm{\gamma}$.
Because of their definition, the entries $\gamma_{i}$ have physical meaning only at the extreme values, that is for $\gamma_{i}=1$ (solid) or $\gamma_{i}=0$ (fluid).
Nevertheless, the problem at hand has been formulated as a continuous optimization problem in Sec. \eqref{sec:optprob} in order to solve it with a computationally efficient gradient-based algorithm.
As a consequence, the variables $\gamma_{i}$ can assume intermediate values between $0$ and $1$.
However, intermediate values should be avoided. This is done by introducing in the final solution appropriate material interpolation functions.
The most popular interpolation scheme in topology optimization is the Solid Isotropic Material with Penalization (SIMP) \citep{Bendsøe1999}. Its main idea is to penalize the intermediate values of a relaxed binary variable definition, implicitly promoting the convergence towards crisp $0-1$ distributions of the design variables (in our case $\bm{\gamma}$).
The Rational Approximation of Material Properties (RAMP) is another material interpolation scheme based on the same idea \citep{Stolpe2001}.
In the optimization problem discussed herein both SIMP and RAMP have been tested.
However, RAMP was chosen for the final problem formulation since it proved to be more effective in leading the optimization algorithm towards final discrete solid-fluid distributions.
In particular, we applied the RAMP scheme to the definition of the material conductivity $k_{i}$ and permeability $\kappa_{i}$ for $i=1,\ldots,N_{ele}$:

\begin{equation}\label{eq:condRAMP}
k_{i}= k_{f} + \frac{\gamma_{i}}{1+q_{c}(1-\gamma_{i})}(k_{s}-k_{f})
\end{equation}
\begin{equation}\label{eq:permRAMP}
\kappa_{i}= \kappa_{s} + \frac{(1-\gamma_{i})}{1+q_{p}\gamma_{i}}(\kappa_{f}-\kappa_{s})
\end{equation}
In Eq. \eqref{eq:condRAMP} and Eq. \eqref{eq:permRAMP}, $q_{c} = 0$ and $q_{p} = 0$ yieldse a linear interpolation of the conductivity and permeability parameters. For increasing values of $q_{c}$ and $q_{p}$, the penalizing effect on the values of $\bm{\gamma}$ between 0 and 1 increases.
Moreover, in Eq. \eqref{eq:condRAMP} $k_{f}$ and $k_{s}$ have fixed values which will be provided in Sec. \ref{sec:numapp}.
In Eq. \eqref{eq:permRAMP}, the definition of $\kappa_{f}$ follows the procedure described in Sec. \ref{sec:tune darcy}, while $\kappa_{s}$ is decreased during the optimization to smoothly converge towards the final optimized designs. More details regarding the values assumed by $\kappa_{s}$ are provided in Sec. \ref{sec:contscheme}. 

\subsection{Continuation scheme}
\label{sec:contscheme}
The material interpolation schemes presented in Eq. \eqref{eq:condRAMP} and Eq. \eqref{eq:permRAMP} are based on few parameters that need to assume specific values in order to produce meaningful interpolations of the material properties. These parameters are $q_{c}$, $q_{p}$, and $\kappa_{s}$. 
Experience showed that it is convenient to start with small but appreciable values of $q_{c}$ and $q_{p}$, and to gradually increase  their value in a step-wise manner in a continuation scheme. In a similar way $\kappa_{s}$ is progressively decreased.
After predefining a maximum allowed number of optimization iterations (e.g. $500$ iterations), the values of these parameters are initialized, and updated at predefined intermediated optimization stages (e.g. every $100$ iterations) according to the following scheme similar to that presented by \citet{Alexandersen2016}:
\begin{equation}\label{eq:contsch1}
\begin{split}
& q_{c} = \left\{0.881, 8.81, 88.1, 88.1, 881.0\right\}\\
& q_{p} = \left\{8.0, 8.0, 8.0,  98.0, 998.0\right\}\\
& \kappa_{s} = \left\{10^{-5}, 10^{-5}, 10^{-5}, 10^{-6}, 10^{-7}\right\}\,m^{2}
\end{split}
\end{equation}
The continuation scheme outlined in \eqref{eq:contsch1} was chosen for our applications because it proved to be effective in gradually converging towards final near-discrete optimized designs.
It should be noted that the optimization problem is highly non-linear and non-convex. The design solutions obtained will be local minima of the problem, and they will strongly depend on the starting point of the optimization analysis, as well as on the particular continuation scheme adopted. 
However, based on the authors' experience, the continuation scheme \eqref{eq:contsch1} gives a good balance between ease of convergence toward the final designs, performance of the optimized designs and modeling accuracy (see \cite{Alexandersen2016} for further discussion).

\subsection{Computational considerations}
\label{sec:compcons}
The optimization and analysis have been implemented in the computational framework developed by \citet{Alexandersen2016}, based on the framework for topology optimization originally presented by \citet{Aage2014} based on PETSc \citep{Balay2018}.
The PETSc framework has been used because it allows for parallelized high performance computing, it provides both linear and non-linear solvers and preconditioners, and the possibility to handle structured meshes.

In particular, the non-linear system of equations is solved with a damped Newton method. The damping coefficient is chosen as the minimizer of the $L^{2}$-norm of the residual vector (i.e. $\textbf{R}(\textbf{s})$) in correspondence with the current intermediate design (optimization) iteration. 
This particular non-linear iterative solver based on the residual minimization combined with a good initial solution guess (the response of the system from the previous optimization iteration) proved to be very robust during the numerical computations performed. 
In the eventuality that the solver fails to converge with the initial solution provided, a ramping scheme is used for the volumetric heat source magnitude (i.e. $Q$).
The stopping criteria for the Newton solver requires a reduction of the $L^{2}$-norm of the residual of $10^{-4}$ relatively to its initial value for the current design iteration.
In practice, this means a high absolute accuracy, since a good initial solution is in general used (the solution from the previous design solution). However, due to the finite precision and approximate solution of the linear systems, the absolute accuracy does not converge to 0.

Most of the computational effort and time is spent on the solution of the linearized system of equations in each Newton iteration. The computational burden originates from the fact that high resolution 3-D design cases are considered. Thus, the linear systems are solved using a Krylov subspace parallelized solver. More details can be found in \cite{Alexandersen2016}. The stopping criteria for the linear solver requires a reduction of the residual of $10^{-5}$ relatively to its initial value.

\section{Benchmark example - heat sink in closed cavity}
\label{sec:numapp}

In this section we present and discuss several numerical results obtained by applying the simplified model to the benchmark example presented by \citet{Alexandersen2016}.
We first test the computational performance of the parallelized framework. The performance is compared with the results of \citet{Alexandersen2016}, where the fluid flow was modeled through the full Navier-Stokes equations.
Then, we discuss the optimization of an academic example of a heat sink. We consider several operational conditions, characterized by different values of the fluid thermal expansion coefficient\footnote{Due to the setting of most parameters to unit size, the thermal expansion coefficient becomes equal to the Grashof number in \cite{Alexandersen2016}.} $\alpha$. 
Lastly, we suggest the use of a two-stage hybrid optimization approach. It consists of two stages, where the two models (simplified and full Navier-Stokes) are used in two sequential stages.
In this way it is possible to obtain final design with a superior performance to those obtained considering the simplified model only, but saving computational efforts.

\subsection{Problem setup}
\begin{figure}\scriptsize
\centering
  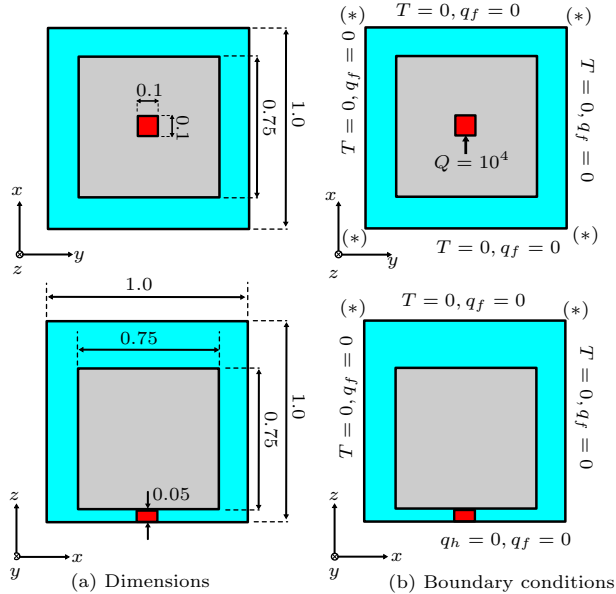\normalsize
\caption{ Representation of the system considered in the numerical applications of Sec. \ref{sec:numapp}. The design domain is grey. The fluid domain is blue and the solid domain with volumetric heat source is red, neither of which are involved in the design. The pressure is zero at at the corners denoted by $(*)$.}
\label{fig:numexdraw}      
\end{figure}
Fig. \ref{fig:numexdraw} shows the problem setup considered in the following examples with details regarding geometry and boundary conditions. The red block represents the heat source (e.g. electronic chip) that generates a volumetric heat $Q=10^{4}\,W/m^{3}$. The design domain is represented by the grey block, which is placed on top of the heat source to allow the cooling fluid to circulate around and beneath it. The external vertical and top walls are kept at a constant temperature $T=0\, K$. The external wall at the bottom is insulated. The dimensions of the outer box are $1 \,m\times 1\,m\times 1\,m$, the design domain dimensions are $0.75\,m\times 0.75\,m \times 0.75\,m$, and the heat source dimensions are $0.1\,m \times 0.1\,m \times 0.05\,m$. 
The boundary conditions and the geometry of the problem are both symmetric. Thus, in the computations we consider a quarter of the original domain with symmetry boundary conditions. The volume fraction is in all examples $5\%$, i.e. $g^{*}=0.05$ in Eq. \eqref{eq:constr}.
The design variables $\gamma_{e}$ are initialized to $0.05$, such that the volume constraint \eqref{eq:constr} is initially satisfied.
The conductivity of the solid material is $k_{s}=100\,W/mK$, while the conductivity of the fluid material is $k_{f}=1\,W/mK$. The remaining parameters are set as follows: $\rho_{0}=1\,kg/m^{3}$, $\mu=1\,Pa\,s$, $c_{p}=1\,J/kgK$.
Lastly, in all the optimization analyses we considered a moving limit of $\Delta \gamma=0.2$. The moving limit defines in each optimization iteration the maximum variable update step from the current solution. The use of moving limits allows for a smooth convergence towards optimized designs, avoiding premature convergence towards undesired poor local minima. 

With regards to the hardware adopted for the computations, the numerical results have been obtained on the exact same cluster as used by \citet{Alexandersen2016}, where each node is composed of two Intel Xeon e5-2680v2 10-core 2.8GHz processors.

\subsection{Computational performance}

To test the computational performance of the state solver, we solved the optimization problem \eqref{eq:optprob} on a fixed mesh for a quarter sub-domain of the problem setup outlined in Fig. \ref{fig:numexdraw}. The mesh resolution considered is $80 \times 80 \times 160$ elements.
The computational performance is shown in Table \ref{tab:jessperformance1}. They have been obtained performing the optimization analyses for $250$ iterations for $\alpha=\{10^{3},10^{6}\}\, K^{-1}$, and with constant values of the penalization parameters: $q_{c}=8.81$ and $q_{p}=8$. Moreover, we considered the values of $\kappa_{f}$ obtained from Eq. \eqref{eq:nusselt5}. The data shows, in both cases, a computational cost that depends almost linearly on the number of processors (CPUs) used. Furthermore, the times show a significant reduction of one order of magnitude compared to the same values reported by \citet{Alexandersen2016}. 
Moreover, compared to the case $\alpha=10^{3}\, K^{-1}$, the computational time for $\alpha=10^{6}\, K^{-1}$ is on average $27.5 \%$ higher.
\begin{table}\centering\small
\caption{Average time $[s]$ required in each optimization iteration by the state solver for a total of $250$ iterations with $\alpha=\{10^{3},10^{6}\}$ $K^{-1}$ and a mesh resolution of $80\times80\times160$ elements.}
\label{tab:jessperformance1}       
\begin{tabular}{lllll}
\hline\noalign{\smallskip}
 & \multicolumn{2}{c}{$\alpha=10^{3}$} & \multicolumn{2}{c}{$\alpha=10^{6}$}\\
\cline{2-3}\cline{4-5}\noalign{\smallskip}
CPUs & Time & Scaling & Time  & Scaling  \\
\hline\noalign{\smallskip}
$16$  & $37.29$ & $1.00$ & $44.77$ & $1.00$\\
$32$  & $17.95$ & $0.48$ & $22.12$ & $0.49$\\
$64$  & $9.09$ & $0.24$ & $11.20$ & $0.25$\\
$160$  & $4.03$ & $0.11$ & $5.13$ & $0.11$\\
$320$  & $2.42$ & $0.06$  & $3.38$ & $0.07$\\
$640$ & $1.25$ & $0.03$  & $1.65$ & $0.04$\\
\hline\noalign{\smallskip}
\end{tabular}
\end{table}

We performed a second study with the purpose of testing the performance of the linear solver (F-GMRES) with respect to the mesh resolution.
The performance was measured in terms of number of linear solver iterations, and we considered both low and high values of the coefficient of thermal expansion $\alpha$. Table \ref{tab:jessperformance3}  lists the number of iterations required by the linear solver averaged for each mesh resolution over $250$, $500$, and $1000$ iterations respectively.
It is seen that the computational cost in general increases with the problem size. That is, the linear solver requires more iterations in average for larger problem sizes. Nevertheless, the actual increase of computational cost is limited, and F-GMRES can be considered an appropriate linear solver for this class of natural convection problems.
Comparing these values to those reported by \citet{Alexandersen2016}, it can be seen that in general the number of linear iterations needed are lower than for the full Navier-Stokes model. This is likely due to the better structure of the linear systems of equations, where both the pressure and temperature problems are Poisson-like equations, leading to further time savings of the proposed model.

\begin{table}\centering\small
\caption{Average iterations required by the linear solver (F-GMRES) over $250$, $500$, $1000$ iterations for $\alpha=\{10^{3},10^{6}\}$ $K^{-1}$ and different mesh resolutions.}
\label{tab:jessperformance3}       
\begin{tabular}{lll}
\hline\noalign{\smallskip}
Mesh & $\alpha=10^{3}$ & $\alpha=10^{6}$  \\
\hline\noalign{\smallskip}
$80\times 80 \times 160$ ($250$ iter)  & $3.8$ & $5.8$ \\
$160\times 160 \times 320$  ($500$ iter)& $ 8.3$ & $6.9$ \\
$320\times 320 \times 640$  ($1000$ iter)& $5.9$ & $11.8$ \\
\hline\noalign{\smallskip}
\end{tabular}
\end{table}

\begin{figure*}
    \centering
    \begin{subfigure}[t]{0.45\textwidth}
        \includegraphics[width=\textwidth]{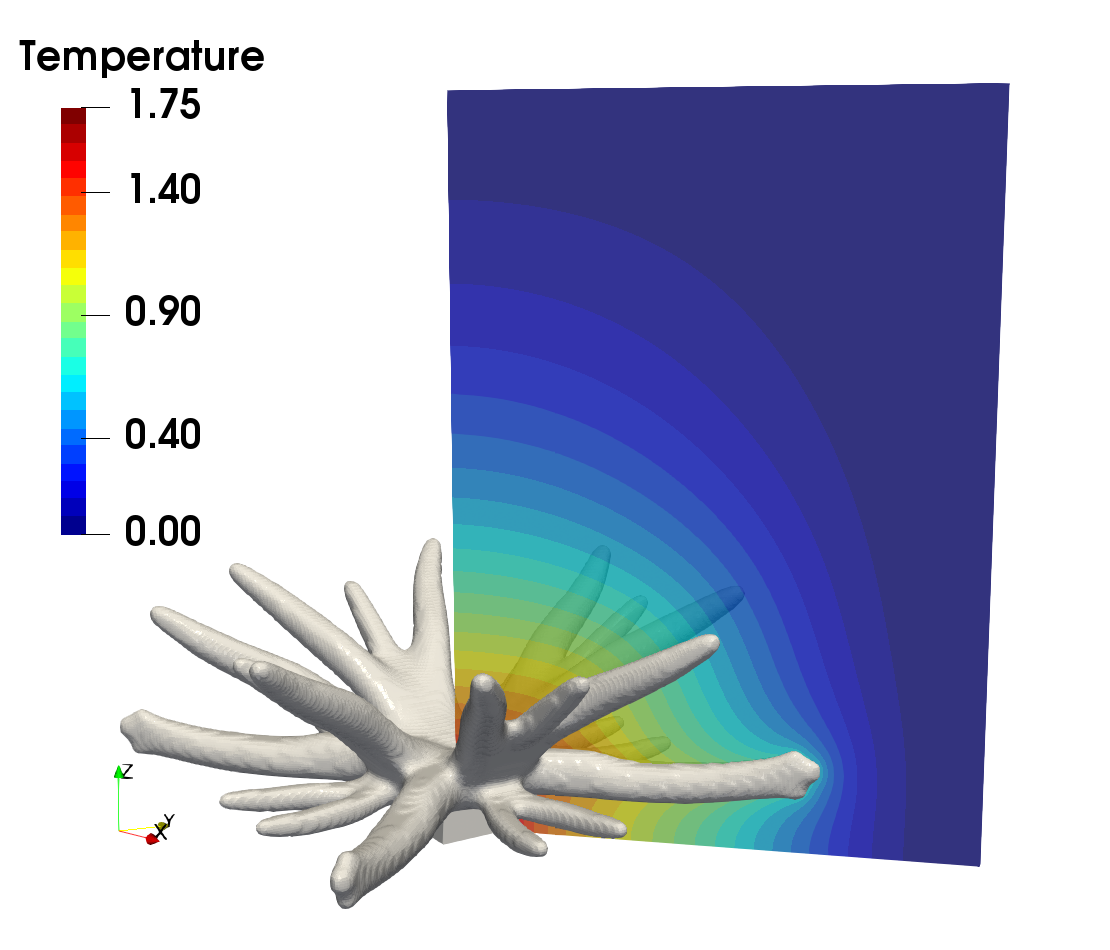}
        \caption{$f=8.53$, $\alpha=10^{3}$, $\kappa_{f}=0.00513$}
        \label{fig:topopt1}
    \end{subfigure}
    \begin{subfigure}[t]{0.45\textwidth}
        \includegraphics[width=\textwidth]{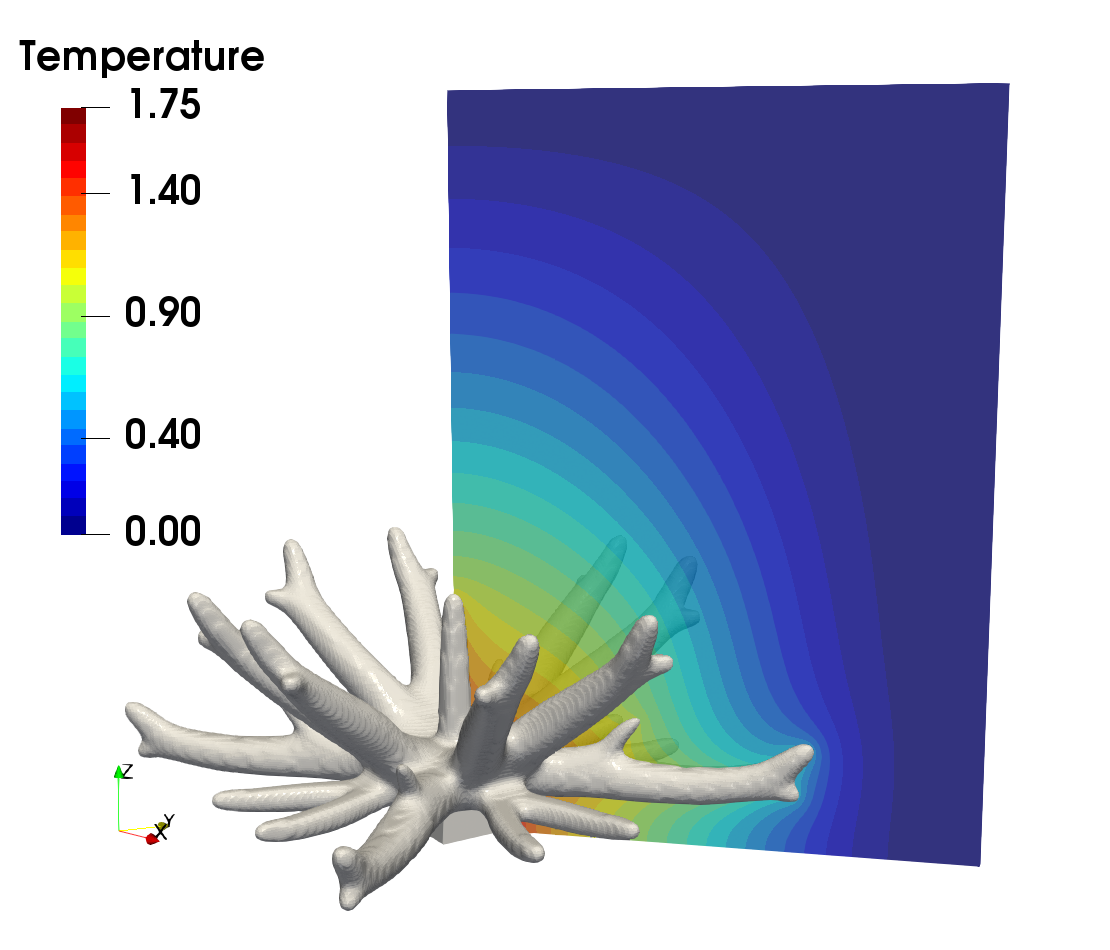}
        \caption{$f=8.28$, $\alpha=10^{4}$, $\kappa_{f}=0.00206$}
        \label{fig:topopt2}
    \end{subfigure}
    \begin{subfigure}[b]{0.45\textwidth}
        \includegraphics[width=\textwidth]{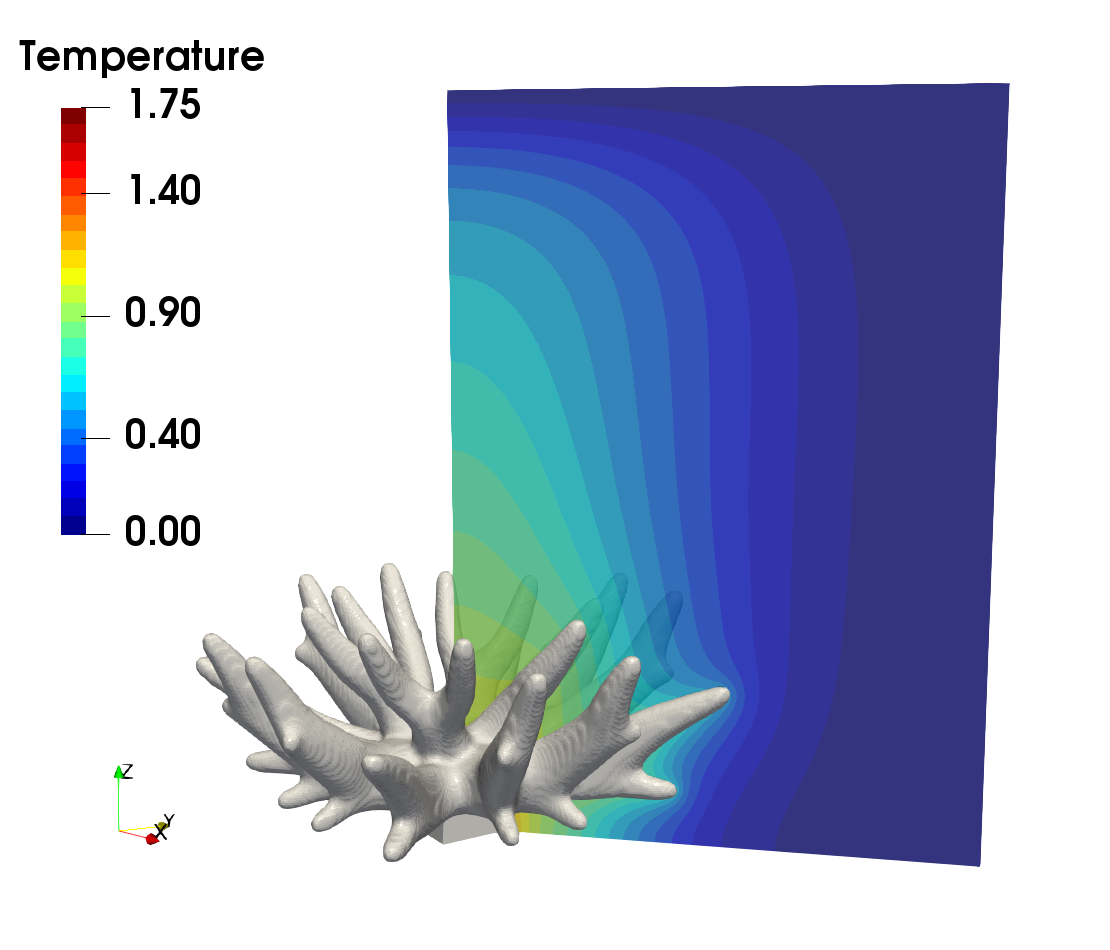}
        \caption{$f=6.67$, $\alpha=10^{5}$, $\kappa_{f}=0.00085$}
        \label{fig:topopt3}
    \end{subfigure}
    \begin{subfigure}[b]{0.45\textwidth}
        \includegraphics[width=\textwidth]{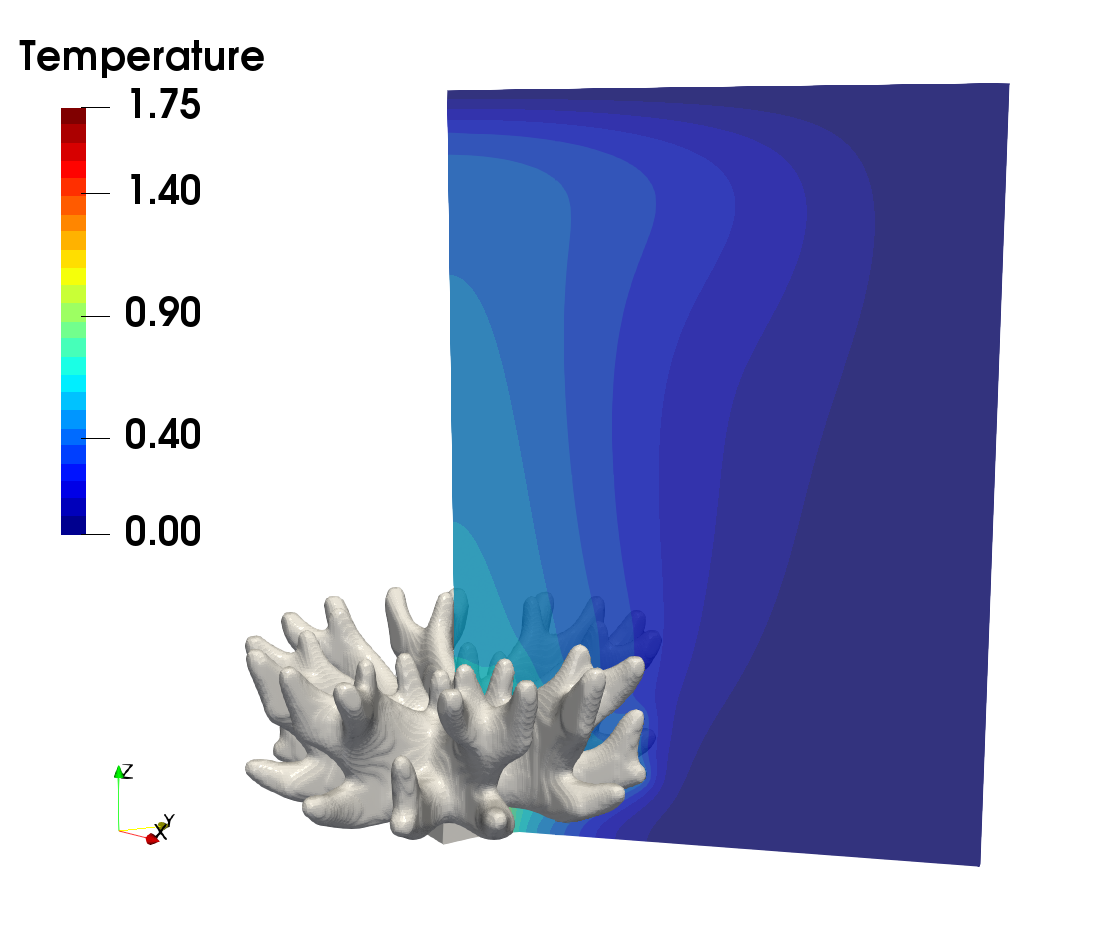}
        \caption{$f=4.81$, $\alpha=10^{6}$, $\kappa_{f}=0.00036$}
        \label{fig:topopt4}
    \end{subfigure}
    \caption{Side view - Temperature $[K]$ disribution in the fluid and optimized topologies for different values of the fluid coefficient of thermal expansion (i.e. $\alpha$ $[K^{-1}]$), and fluid permability (i.e. $\kappa_{f}$ $[m^{2}]$). Results obtained with a mesh resolution of $160\times 160\times 320$ elements, and performing $500$ optimization iterations with $500$ CPUs.}\label{fig:optimdesign}
\end{figure*}

\subsection{Design for different thermal expansion coefficients of the fluid}
\label{sec:numexdiffalpha}
In this section, we present the results obtained considering different values of the coefficient of thermal expansion, $\alpha$. With these numerical examples we intend to provide an insight into the effect of the governing parameter for the fluid-thermal coupling on the final optimized designs.
The results have been obtained considering a computational mesh of $160\times 160 \times 320$ elements, which resulted in a total of $8,192,000$ elements and $16,384,000$ degrees of freedom.
The optimization analyses discussed in this section have been performed with $500$ CPUs for $500$ iterations. 
The design domain consists of $3,456,000$ elements and the filter radius was set to 2.5 times the element size, i.e. $0.003125 \times 2.5 = 0.0078125 \, m$.
Fig. \ref{fig:optimdesign} shows the optimized designs obtained for different values of $\alpha$.
Table \ref{tab:topoptperformance} lists the time required for the optimization analyses, and the final values of the objective function.
Once again comparing to the computational times reported by \citet{Alexandersen2016}, significantly shorter time ($27\%-55\%$) is needed using only 500 cores in contrast to 1280. In the unit of core-hours, this yields a total reduction of the computational time to $10-20\%$.

\begin{figure*}
    \centering
    \begin{subfigure}[b]{0.475\textwidth}\centering
        \includegraphics[width=\textwidth]{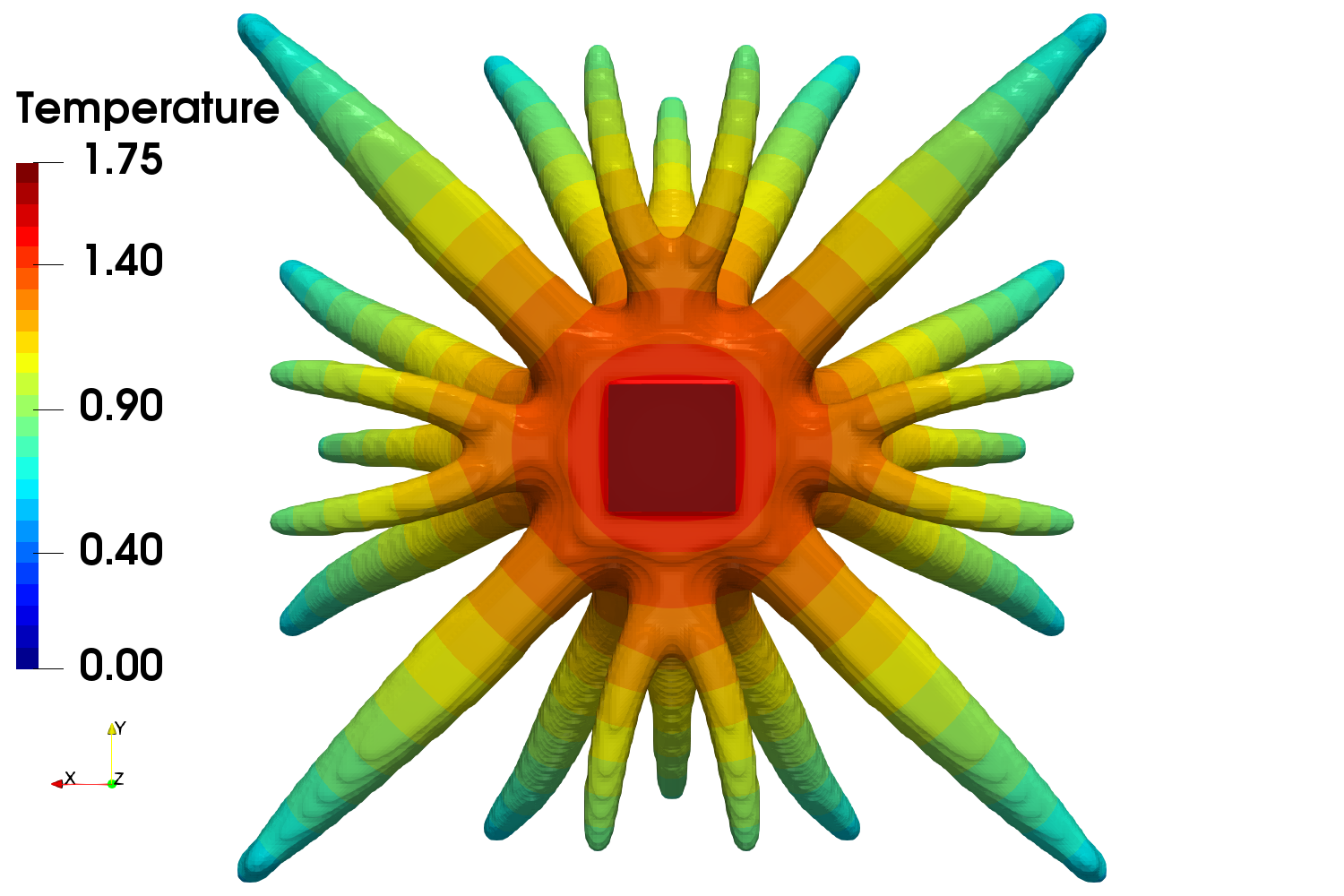}
        \caption{$f=8.53$, $\alpha=10^{3}$, $\kappa_{f}=0.00513$}
        \label{fig:topoptbottom1}
    \end{subfigure}
    ~ 
    \begin{subfigure}[b]{0.475\textwidth}\centering
        \includegraphics[width=\textwidth]{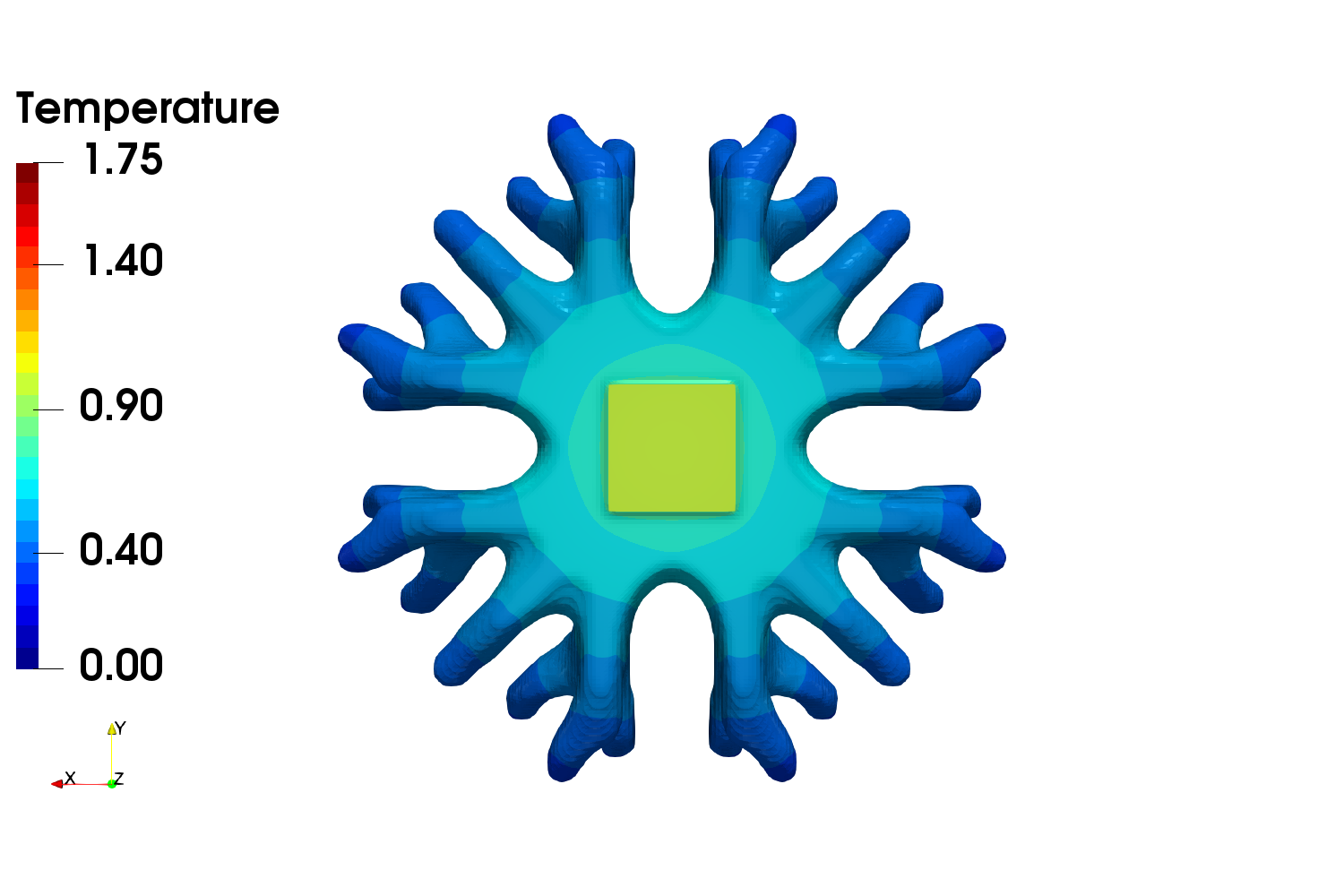}
        \caption{$f=4.81$, $\alpha=10^{6}$, $\kappa_{f}=0.00036$}
        \label{fig:topoptbottom2}
    \end{subfigure}
    \caption{Bottom view - Temperature $[K]$ distribution in the optimized designs for $\alpha=\{10^{3},10^{6}\}$ $K^{-1}$ and $\kappa _{f}=\{0.00513,0.00036\}$ $m^{2}$. Results obtained with a mesh resolution of $160\times 160 \times 320$ elements, and performing $500$ optimization iterations with $500$ CPUs.}\label{fig:topoptbottom}
\end{figure*}

The optimized designs have been visualized in ParaView \citep{Ayachit2018}, where the densities have been thresholded at $\gamma = 0.9$.
Strong similarities with the results presented by \citet{Alexandersen2016} verify the assumptions behind the simplified fluid model.
First, the optimized topologies are characterized by ``thermal tree'' shapes that carry the heat generated away from the source to cooler areas in the surrounding fluid.
Another similarity is the tendency of the optimized topologies to contract and to increase the number of branches as the value of $\alpha$ increases. This can be explained by the shift from a conduction/diffusion dominated problem, to one dominated by convection. For lower values of $\alpha$ the heat is mostly dissipated by conduction, and this results in longer branches that carry the load to the cool areas of the fluid close to the boundaries. For higher values of $\alpha$, the heat is instead dissipated mostly through convection. To this end, the optimizer identifies topologies with an increased surface area at the fluid-solid interface where the heat is exchanged with the fluid, and carried away by the flow. 
Fig. \ref{fig:topoptbottom} shows the optimized designs for $\alpha=10^{3}$ $K^{-1}$ and $\alpha=10^{6}$ $K^{-1}$ viewed from below.
Notice the tendency to reach more peripheral and cold areas of the fluid in the first case, and a more contracted shape in the second case.
The performance of the obtained designs has been evaluated for all the operating conditions considered, i.e. $\alpha$ values.
The results are listed in Table  \ref{tab:petscperformance}.
It can be observed that each design performs at its best in the operating condition for which it is optimized.

\begin{table}\centering\small
\caption{Computational time $[hours]$ required by the optimization analyses and final values of the objective function for different $\alpha$ $[K^{-1}]$ values. The results refer to the optimized designs shown in Fig. \ref{fig:optimdesign}.}
\label{tab:topoptperformance}       
\begin{tabular}{lllll}
\hline\noalign{\smallskip}
$\alpha$ & $10^{3}$& $10^{4}$& $10^{5}$ & $10^{6}$  \\
\hline\noalign{\smallskip}
Time  & $5.5$  & $2.8$ & $3.5$ & $4.2$ \\
$f$ & $8.53$ & $8.28$ & $6.67$ & $4.81$ \\
\hline\noalign{\smallskip}
\end{tabular}
\end{table}

\begin{table}\centering\small
\caption{Cross-check of the objective function for the verification of the designs of Fig. \ref{fig:optimdesign} considering the simplified flow model. Text coloring between blue (cold) and red (hot) is scaled according to the designs with minimum and maximum performance for each analysis case (i.e. $\alpha$ value $[K^{-1}]$). Bold numbers highlight the minimum value of the objective function for a given analysis case.}
\label{tab:petscperformance}       
\begin{tabular}{lllll}
\hline\noalign{\smallskip}
 & \multicolumn{4}{c}{Optimization $\alpha$}  \\
\cline{2-5}\noalign{\smallskip}
Analysis $\alpha$ & $10^{3}$& $10^{4}$&$10^{5}$ &$10^{6}$\\
\hline\noalign{\smallskip}
$10^{3}$  & $\textcolor{mycolor51}{\bm{8.53}}$ & $\textcolor{mycolor52}{8.56}$ & $\textcolor{mycolor53}{9.05}$& $\textcolor{mycolor54}{9.76}$\\
$10^{4}$  & $\textcolor{mycolor61}{8.30}$ & $\textcolor{mycolor62}{\bm{8.28}}$& $\textcolor{mycolor63}{8.59}$& $\textcolor{mycolor64}{9.12}$  \\
$10^{5}$ & $\textcolor{mycolor71}{6.87}$ & $\textcolor{mycolor72}{6.76}$ & $\textcolor{mycolor73}{\bm{6.67}}$& $\textcolor{mycolor74}{6.76}$ \\
$10^{6}$ & $\textcolor{mycolor81}{5.28}$ & $\textcolor{mycolor82}{5.17}$ & $\textcolor{mycolor83}{4.93}$& $\textcolor{mycolor84}{\bm{4.81}}$ \\
\hline\noalign{\smallskip}
\end{tabular}
\end{table}

\begin{figure*}
    \centering
    \begin{subfigure}[t]{0.425\textwidth}
        \includegraphics[width=\textwidth]{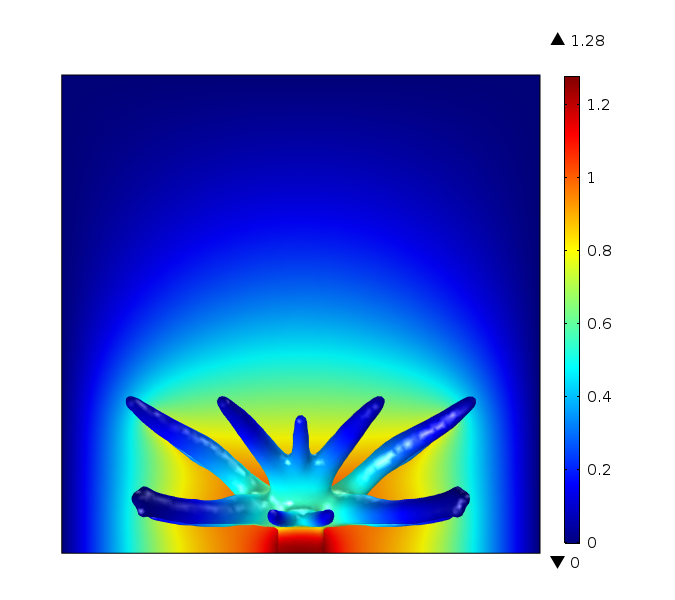}
        \caption{Temperature ($\alpha=10^{3}$)}
        \label{fig:comsol1}
    \end{subfigure}
    \begin{subfigure}[t]{0.425\textwidth}
        \includegraphics[width=\textwidth]{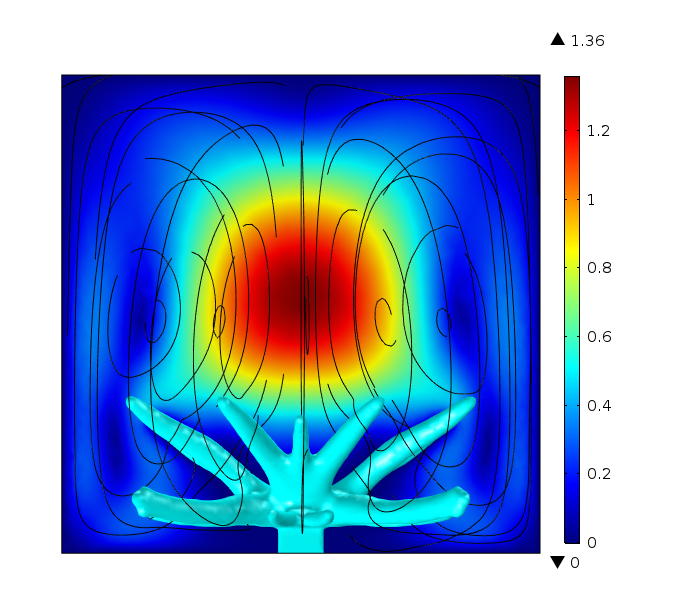}
        \caption{Velocity magnitude and streamlines ($\alpha=10^{3}$)}
        \label{fig:comsol2}
    \end{subfigure}
    \begin{subfigure}[b]{0.425\textwidth}
        \includegraphics[width=\textwidth]{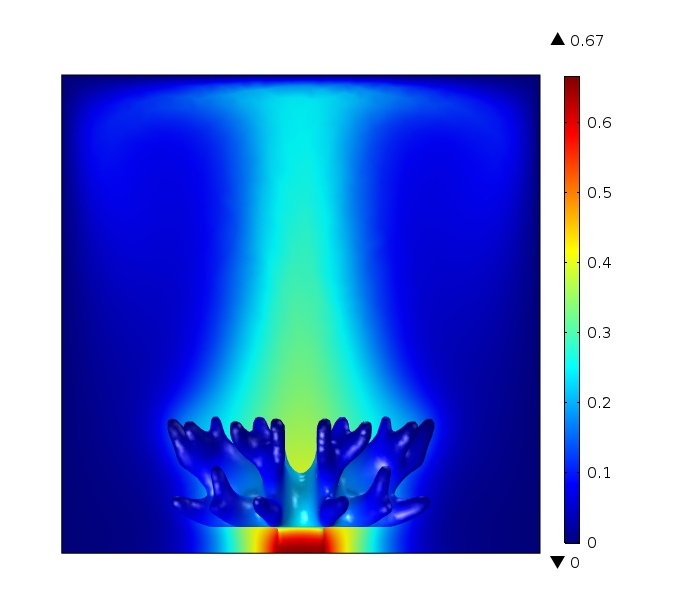}
        \caption{Temperature ($\alpha=10^{6}$)}
        \label{fig:comsol3}
    \end{subfigure}
    \begin{subfigure}[b]{0.425\textwidth}
        \includegraphics[width=\textwidth]{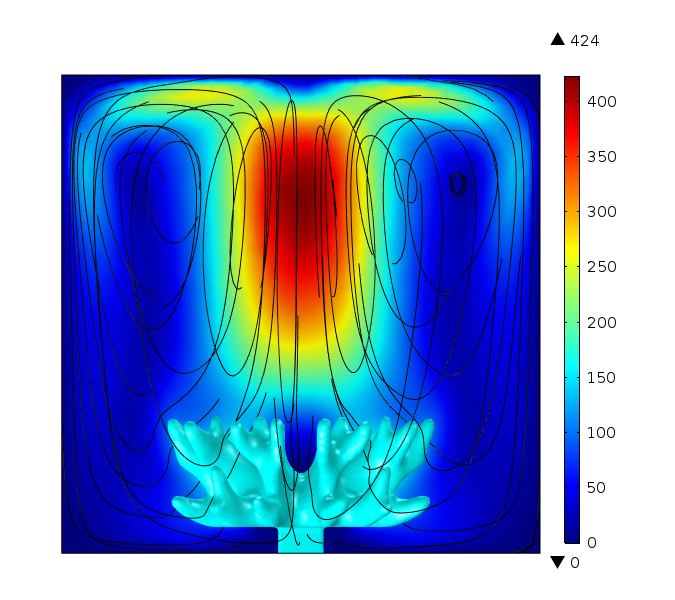}
        \caption{Velocity magnitude and streamlines ($\alpha=10^{6}$)}
        \label{fig:comsol4}
    \end{subfigure}
    \caption{Verification in COMSOL of the optimized designs presented in Sec. \ref{sec:numexdiffalpha} for $\alpha=\{10^{3},10^{6}\}$ $K^{-1}$ considering incompressible steay-state laminar Navier-Stokes flow. Temperature $[K]$ and velocity magnitude $[m/s]$ plots.
    }\label{fig:optimdesigncomsol}
\end{figure*}

To verify the performance of the optimised designs, the designs have been thresholded at $\gamma=0.9$ and exported from ParaView as smooth surfaces. They have then been analyzed using a body-fitted mesh in COMSOL Multiphysics 5.3 \citep{COMSOL}, considering a high-fidelity conjugate heat transfer model. 
For modeling the designs in COMSOL, we considered the same parameter setting used in the optimization.
Each of the optimized designs displayed in Fig. \ref{fig:optimdesign} was tested for all the values of $\alpha$. 
The intention was to check whether each design would have performed better than the others for the specific value of $\alpha$ for which it was optimized for, also considering the high-fidelity model.
\begin{table}\centering\small
\caption{Cross-check of the objective function for the verification of the designs of Fig. \ref{fig:optimdesign} using COMSOL. Text coloring between blue (cold) and red (hot) is scaled according to the designs with minimum and maximum performance for each analysis case (i.e. $\alpha$ value $[K^{-1}]$). Bold numbers highlight the minimum value of the objective function for a given analysis case.}
\label{tab:comsolperformance}       
\begin{tabular}{lllll}
\hline\noalign{\smallskip}
 & \multicolumn{4}{c}{Optimization $\alpha$}  \\
\cline{2-5}\noalign{\smallskip}
Analysis $\alpha$ & $10^{3}$& $10^{4}$&$10^{5}$ &$10^{6}$\\
\hline\noalign{\smallskip}
$10^{3}$  & $\textcolor{mycolor11}{\bm{5.55}}$ & $\textcolor{mycolor12}{5.64}$ & $\textcolor{mycolor13}{6.31}$& $\textcolor{mycolor14}{7.11}$\\
$10^{4}$  & $\textcolor{mycolor21}{\bm{5.27}}$ & $\textcolor{mycolor22}{5.28}$& $\textcolor{mycolor23}{5.77}$& $\textcolor{mycolor24}{6.37}$  \\
$10^{5}$ & $\textcolor{mycolor31}{4.02}$ & $\textcolor{mycolor32}{\bm{3.94}}$ & $\textcolor{mycolor33}{3.99}$& $\textcolor{mycolor34}{4.24}$ \\
$10^{6}$ & $\textcolor{mycolor41}{2.92}$ & $\textcolor{mycolor42}{2.88}$ & $\textcolor{mycolor43}{\bm{2.75}}$& $\textcolor{mycolor44}{2.79}$ \\
\hline\noalign{\smallskip}
\end{tabular}
\end{table}
The results obtained from the verification in COMSOL are shown in Table \ref{tab:comsolperformance}. Each column contains the performances of each of the optimized designs. 
Fig. \ref{fig:optimdesigncomsol} shows the performances in terms of temperature and velocity magnitude obtained in COMSOL of the deigns optimized for $\alpha=\{10^{3},10^{6}\}$ $K^{-1}$.
The results in Table \ref{tab:comsolperformance} show that only in the first case (i.e. for $\alpha=10^{3}$ $K^{-1}$) the design outperforms the others for the specific flow condition for which it is optimized. However, for the other flow conditions, the designs have a performance quite close to the best one for that specific condition.
Some perform better for flow conditions that are different from those considered for their optimization. 
Clearly, the topology optimization approach discussed herein based on the Darcy flow model should not be expected to identify optimized designs also for the case of Navier-Stokes flow.
Rather, it should be seen as a approach that reduces the computational cost significantly, and that provides unintuitive designs that would not have been possible otherwise to identify with even simpler models. 

\subsection{Hybrid optimization approach}
\label{sec:hybridnatconv}
In the previous section, it has been possible to observe that the designs obtained considering the simplified fluid flow model had a small performance loss when tested with the full Navier-Stokes fluid flow model.
This was an expected consequence of the fact that the models considered for optimization and for verification were different. 
In this section, we present an additional application where we combine the use of the simplified fluid flow model discussed herein with the full Navier-Stokes model. The goal is to obtain optimized designs that possessed a better performance than those optimized considering the simplified flow model only, if tested with the full Navier-Stokes fluid flow model.

\begin{figure*}
    \centering
    \begin{subfigure}[t]{0.44\textwidth}
        \includegraphics[width=\textwidth]{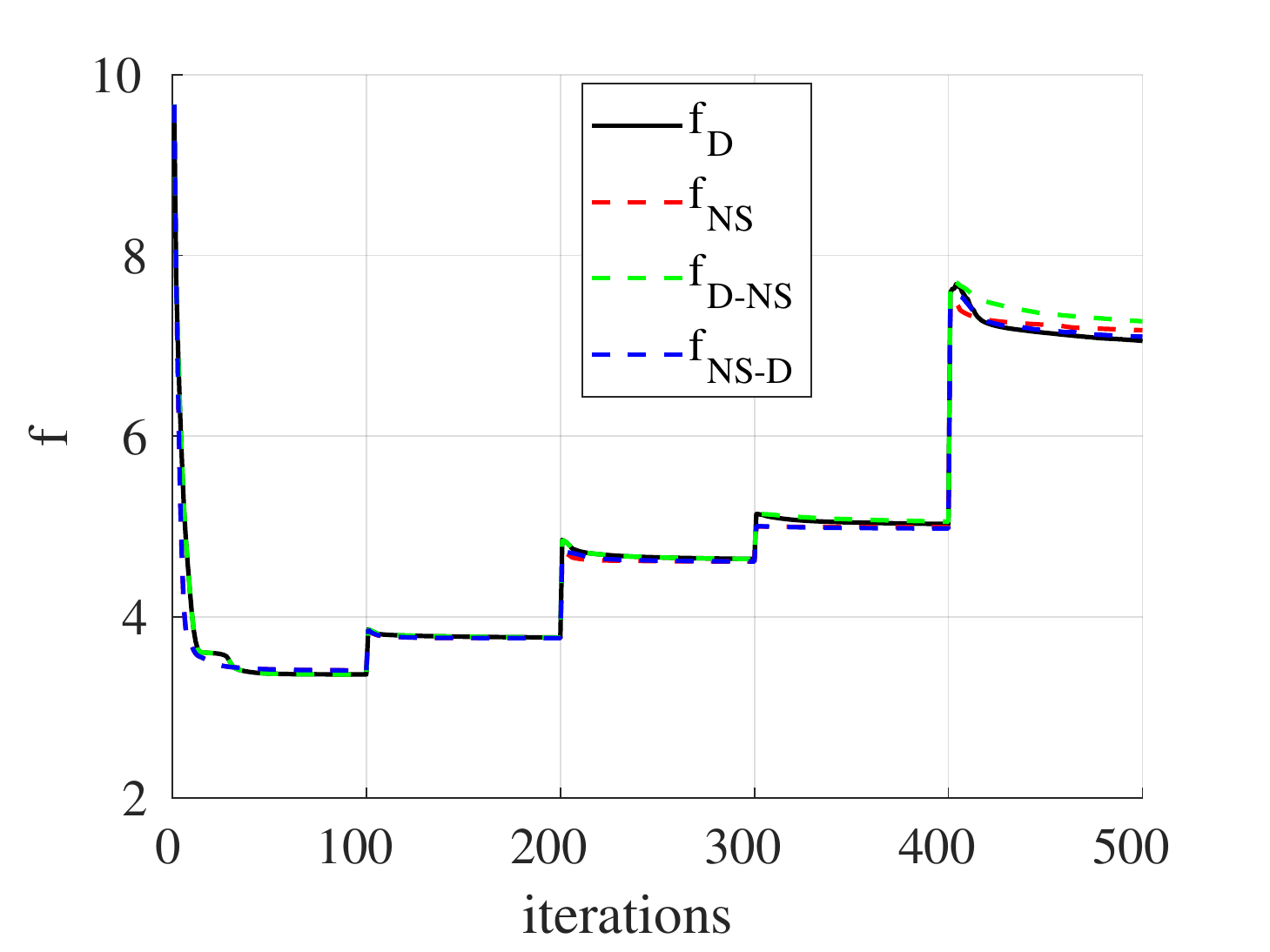}
        \caption{Performance measured with Darcy flow }
        \label{fig:hybrid1}
    \end{subfigure}
    ~ 
    \begin{subfigure}[t]{0.44\textwidth}
        \includegraphics[width=\textwidth]{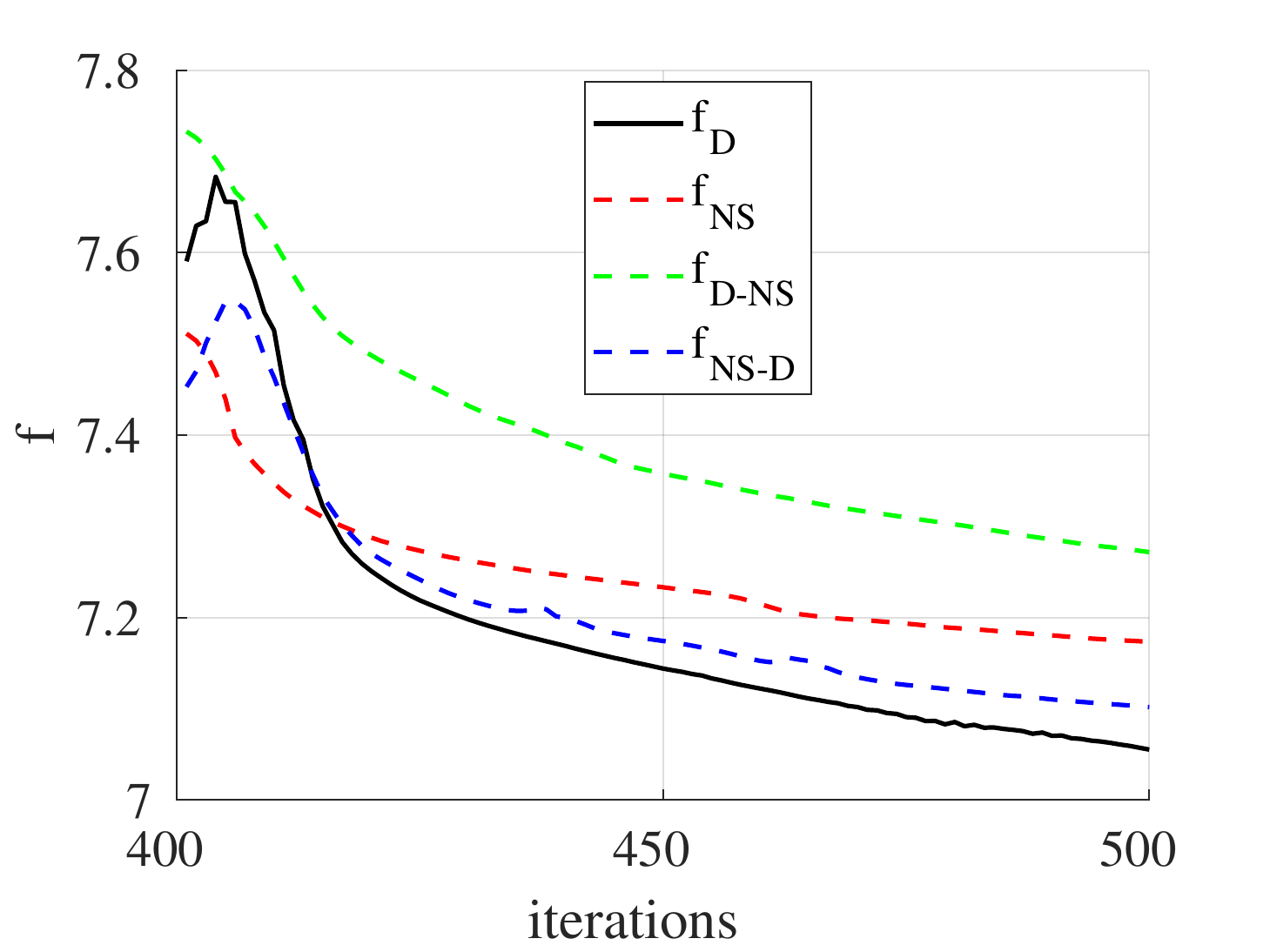}
        \caption{Performance measured with Darcy flow}
        \label{fig:hybrid2}
    \end{subfigure}
    \begin{subfigure}[b]{0.44\textwidth}
        \includegraphics[width=\textwidth]{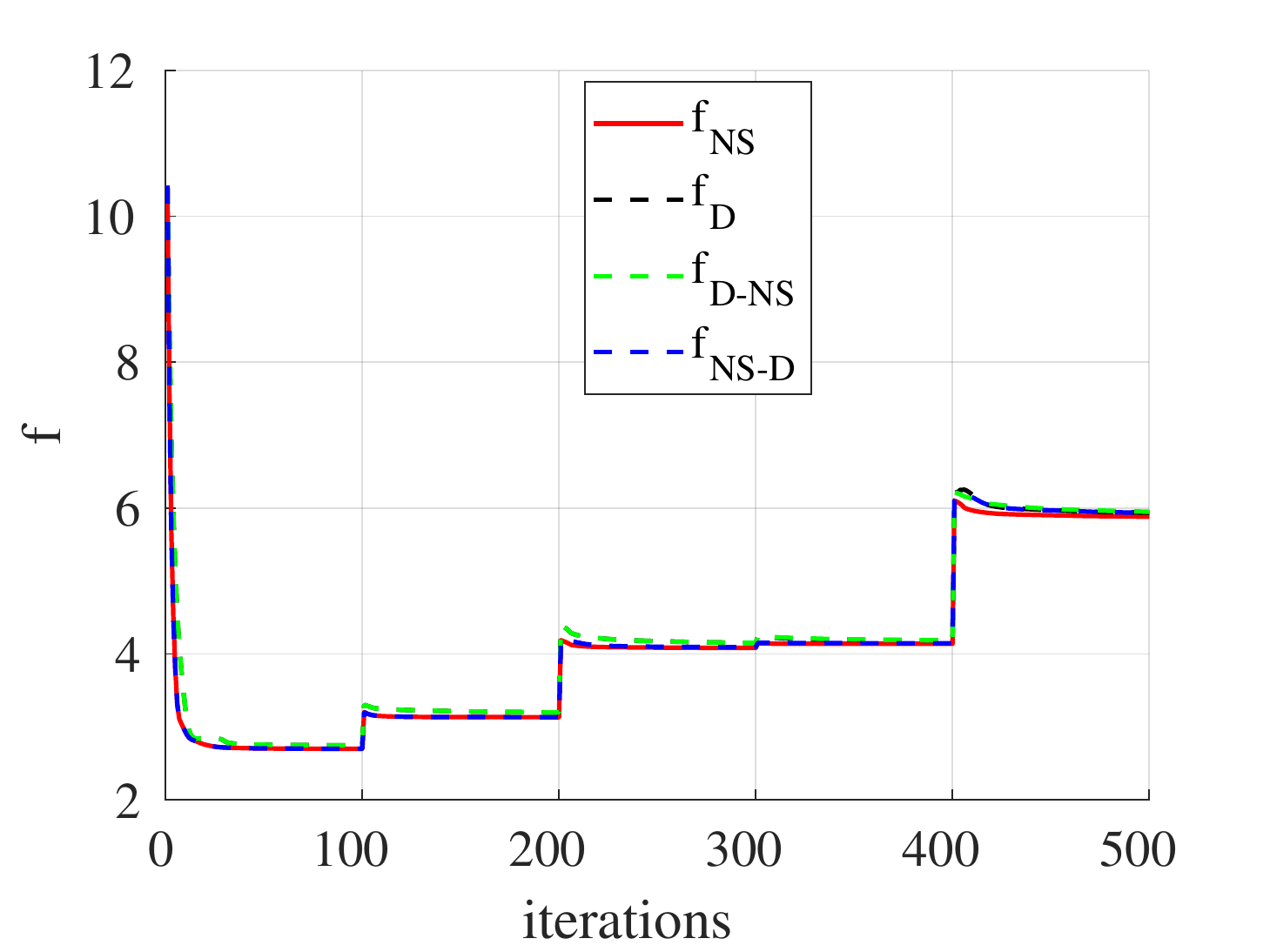}
        \caption{Performance measured with Navier-Stokes flow}
        \label{fig:hybrid3}
    \end{subfigure}
    \begin{subfigure}[b]{0.44\textwidth}
        \includegraphics[width=\textwidth]{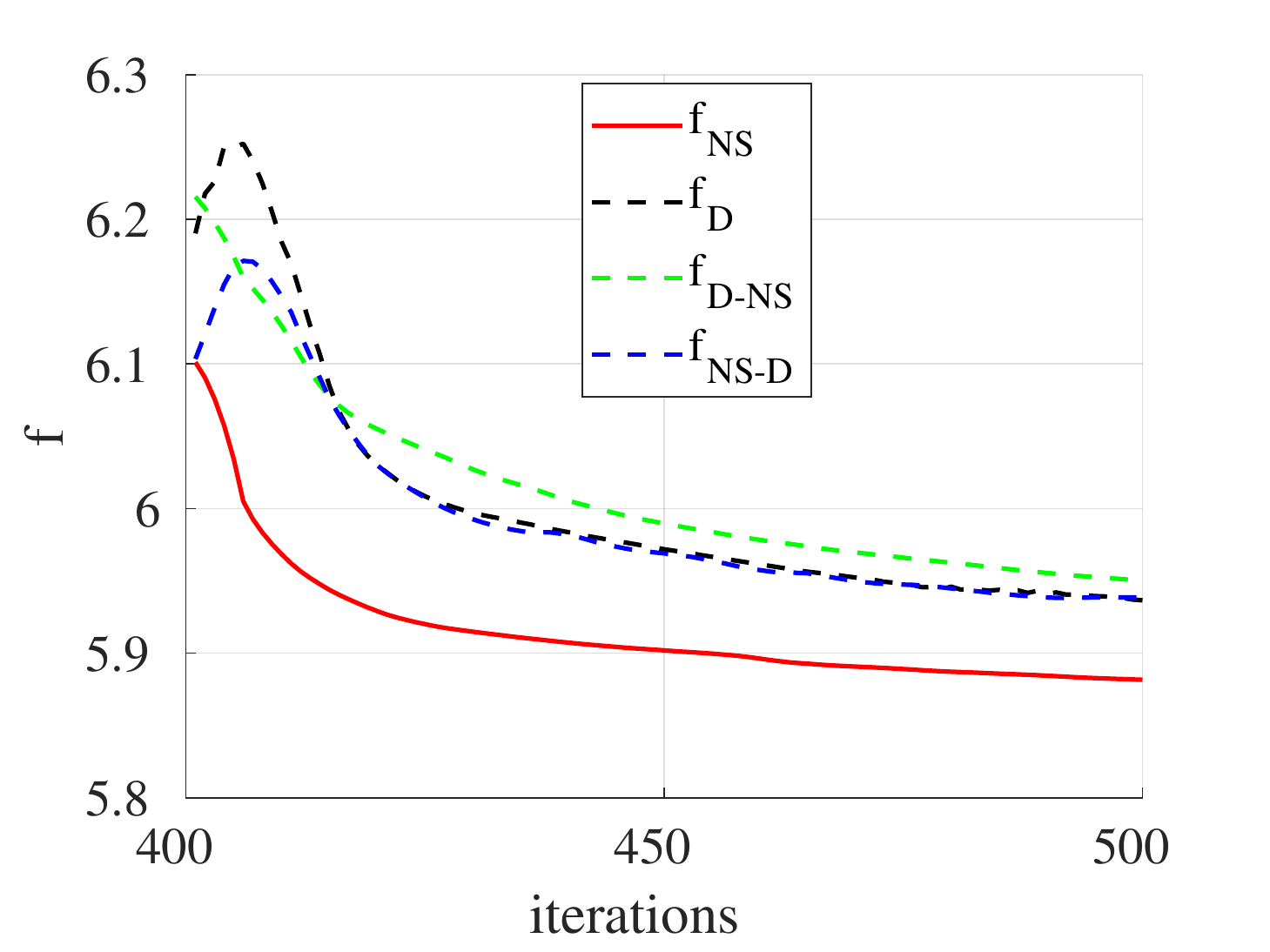}
        \caption{Performance measured with Navier-Stokes flow}
        \label{fig:hybrid4}
    \end{subfigure}
    \caption{Comparison of the performances of the different optimization strategies outlined in Table \ref{tab:hybridapproaches}. (\ref{fig:hybrid1}) and (\ref{fig:hybrid2}): $f$ evaluated considering the Darcy flow model; (\ref{fig:hybrid3}) and (\ref{fig:hybrid4}): $f$ evaluated considering the Navier-Stokes flow model. Design case: heat sink in closed box (Fig. \ref{fig:numexdraw}) with $\alpha=10^{6}$ $K^{-1}$, and with a resolution of $80 \times 80 \times 160$ elements.}\label{fig:hyb_mod_compare}
\end{figure*}

With the intention of maintaining the overall computational cost required within reasonable limits, we considered a hybrid optimization approach, where both the Darcy and the Navier-Stokes flow models were considered in two sequential stages.
This requires switching the fluid model considered at a predefined intermediate optimization step.
In order to identify a suitable optimization strategy based on a hybrid approach, we performed a study where we compared different approaches.
We considered a single design case as described in Fig. \ref{fig:numexdraw} for $\alpha=10^{6}$ $K^{-1}$ and with a mesh resolution of $80\times 80 \times 160$ elements. We performed the optimization considering different fluid flow models used for the evaluation of the performance (i.e thermal compliance, $f$):
optimization based on the Darcy model, $f_{D}$; optimization based on the Navier-Stokes model, $f_{NS}$; optimization based on the Darcy model followed by the Navier-Stokes model, $f_{D-NS}$; and optimization based on the Navier-Stokes model followed by the Darcy model, $f_{NS-D}$. In Table \ref{tab:hybridapproaches} we provide a description of the different models considered during the optimization according to the different approaches.
\begin{table}[hbt]\centering\small
\caption{Different strategies adopted to model the natural convection during the optimization in Sec. \ref{sec:hybridnatconv} ($D$: Darcy; $NS$: Navier-Stokes).}
\label{tab:hybridapproaches}       
\begin{tabular}{lllll}
\hline\noalign{\smallskip}
  & \multicolumn{4}{c}{Model for optimization}  \\
\cline{2-5}\noalign{\smallskip}
Optimization\\iteration & $D$& $NS$ & $D$-$NS$ &$NS$-$D$  \\
\hline\noalign{\smallskip}
$1$-$100$ & $D$& $NS$ & $D$ & $NS$ \\
$101$-$200$ & $D$ & $NS$ & $D$ & $NS$ \\
$201$-$300$ & $D$ & $NS$ & $D$ & $D$ \\
$301$-$400$ & $D$ & $NS$ & $NS$& $D$\\
$401$-$500$ & $D$ & $NS$ & $NS$& $D$\\
\hline\noalign{\smallskip}
\end{tabular}
\end{table}
In Fig. \ref{fig:hyb_mod_compare}, we compare the performance of the different approaches during the optimization analyses.
The performances of the four approaches outlined in Table \ref{tab:hybridapproaches} are compared in terms of thermal compliance (Eq. \eqref{eq:objfunc}), considering the simplified (i.e. Darcy) and full (i.e. Navier-Stokes) fluid flow models.
Fig. \ref{fig:hybrid2} shows the evolution of the performance of the different approaches during the last step of the continuation scheme (i.e. last $100$ optimization iterations). The performance is compared considering the Darcy flow model. As expected, the design optimized for the Darcy model (i.e. $D$) has the best final performance. 
Among the hybrid approaches (i.e. $D$-$NS$, $NS$-$D$), $NS$-$D$ identified an optimized design with a better performance.
Similarly, Fig. \ref{fig:hybrid4} shows the performances calculated considering the Navier-Stokes flow model. Also in this case, the design optimized for the same model used for verification has the best performance (i.e. $NS$), and the approach based on the hybrid approach $NS$-$D$ identified a design with better performance.
 
Based on the results shown in Fig. \ref{fig:hyb_mod_compare}, we decided to use the $NS$-$D$ hybrid approach for further optimization.
In this way, the more accurate, but computationally more expensive, Navier-Stokes flow model is considered only in the initial stages of the optimization-based design. 
As a result, initially the algorithm is expected to converge towards a local optimal design characterized by a better performance if analyzed considering a Navier-Stokes fluid flow. In the second phase, we significantly reduce the computational effort (and time) required by the optimization analysis by considering the simplified flow model. 
In this stage, we finalize the design initially identified converging towards a final near-discrete optimized topology.

\begin{figure*}
    \centering
    \begin{subfigure}[b]{0.45\textwidth}
        \includegraphics[width=\textwidth]{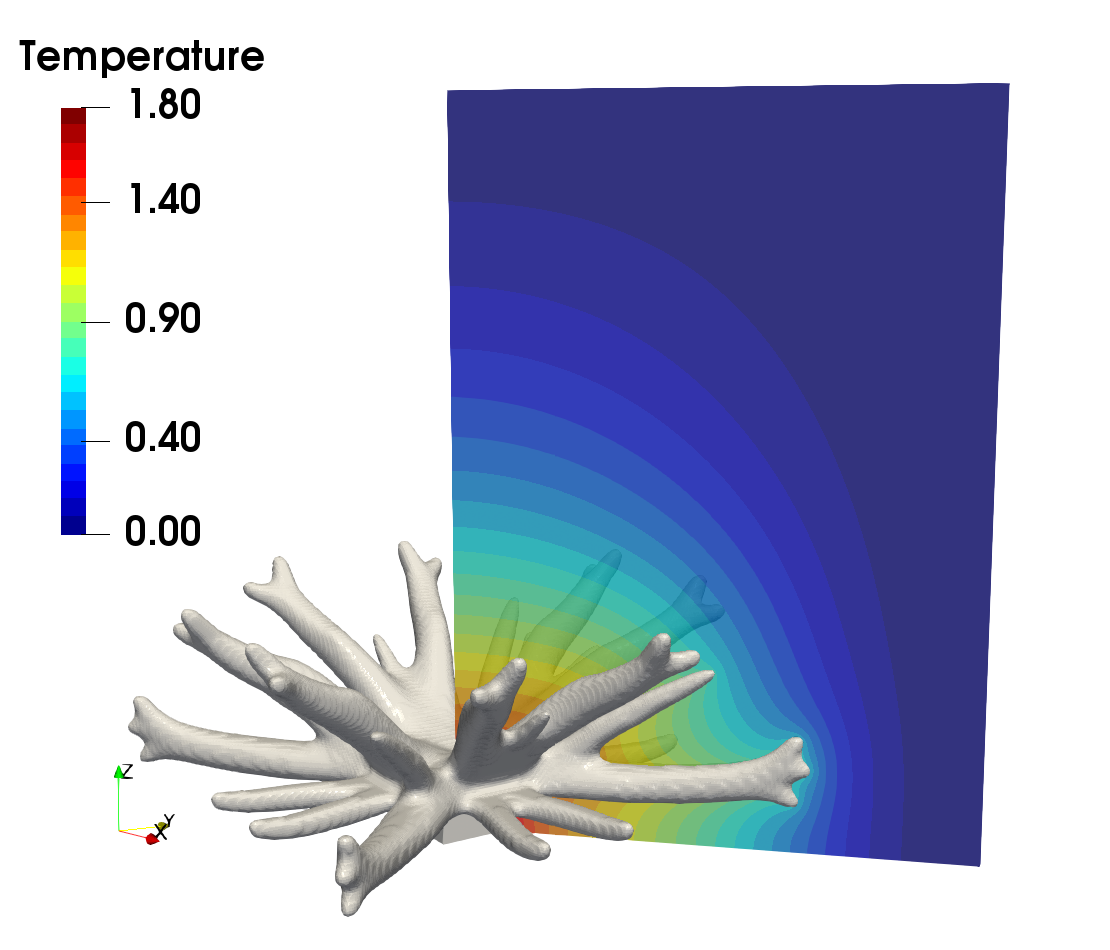}
        \caption{$f=8.76$, $\alpha=10^{3}K^{-1}$}
        \label{fig:gull}
    \end{subfigure}
    ~ 
    \begin{subfigure}[b]{0.45\textwidth}
        \includegraphics[width=\textwidth]{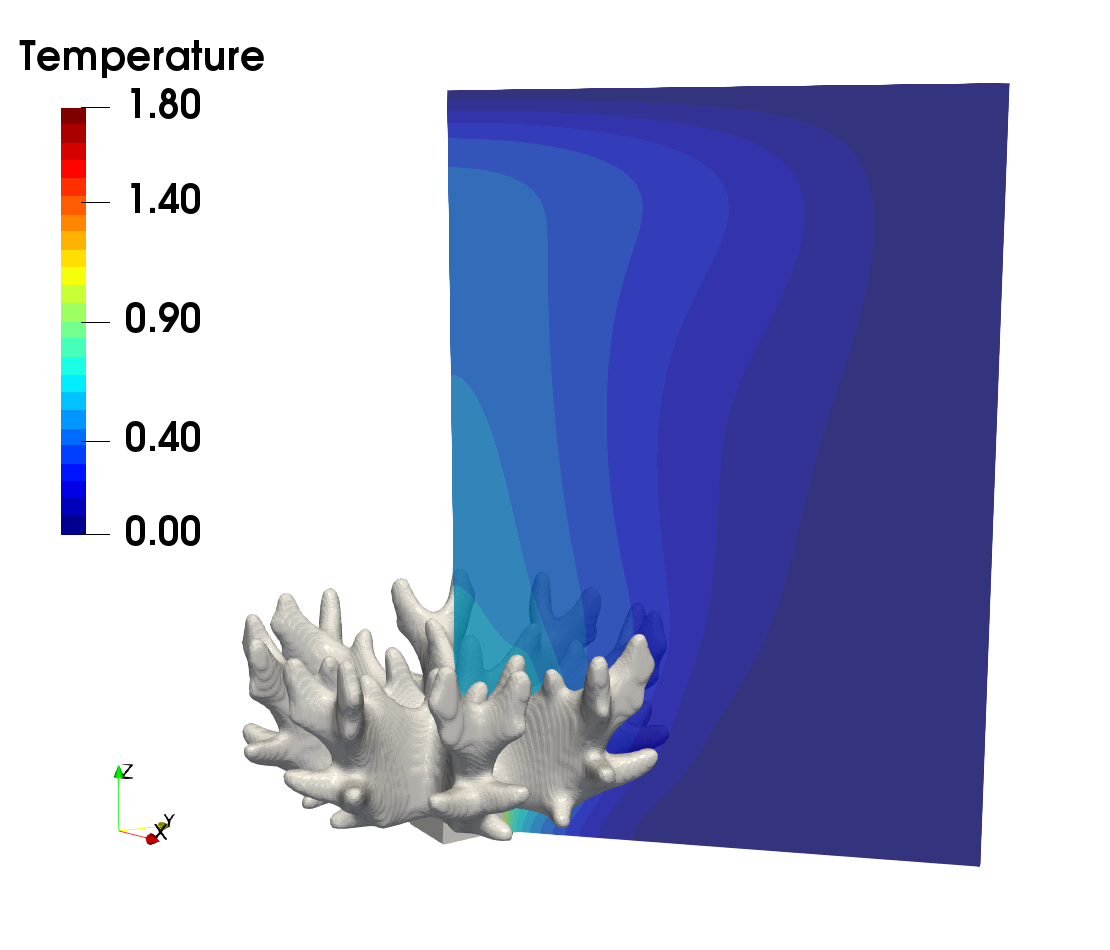}
        \caption{$f=4.78$, $\alpha=10^{6}K^{-1}$}
        \label{fig:tiger}
    \end{subfigure}
    \caption{Side view - Temperature $[K]$ distribution in the fluid and optimized topologies for $\alpha=\{10^{3},10^{6}\}$ $K^{-1}$. Results obtained with a mesh resolution of $160\times 160 \times 320$ elements, and performing $200$ optimization iterations considering a Navier-Stokes fluid flow followed by $300$ optimization iterations considering a Darcy fluid flow.}\label{fig:NS+D1}
\end{figure*}

\begin{figure*}
    \centering
    \begin{subfigure}[b]{0.475\textwidth}
        \includegraphics[width=\textwidth]{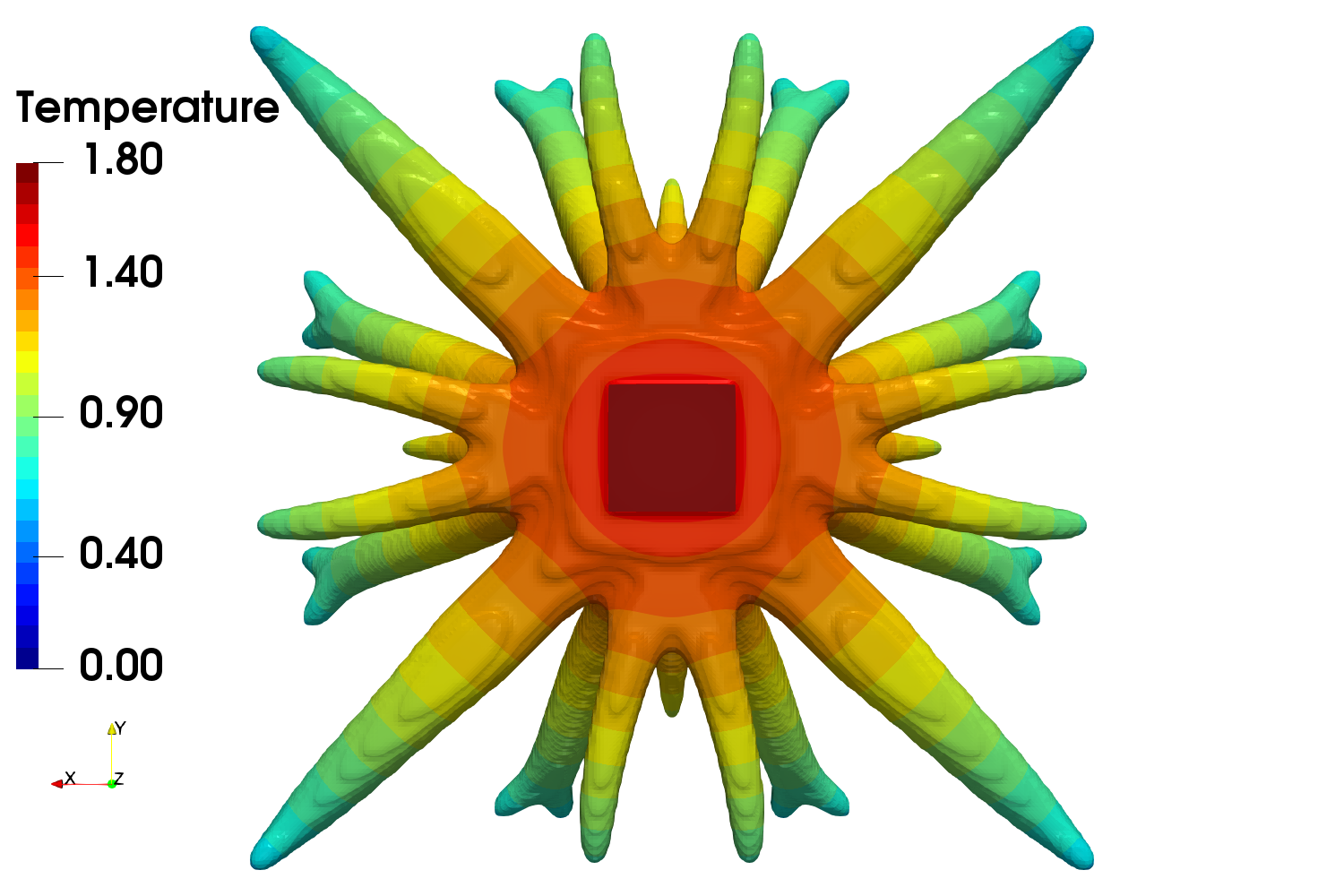}
        \caption{$\alpha=10^{3}K^{-1}$}
        \label{fig:shuligull2}
    \end{subfigure}
    ~ 
    \begin{subfigure}[b]{0.475\textwidth}
        \includegraphics[width=\textwidth]{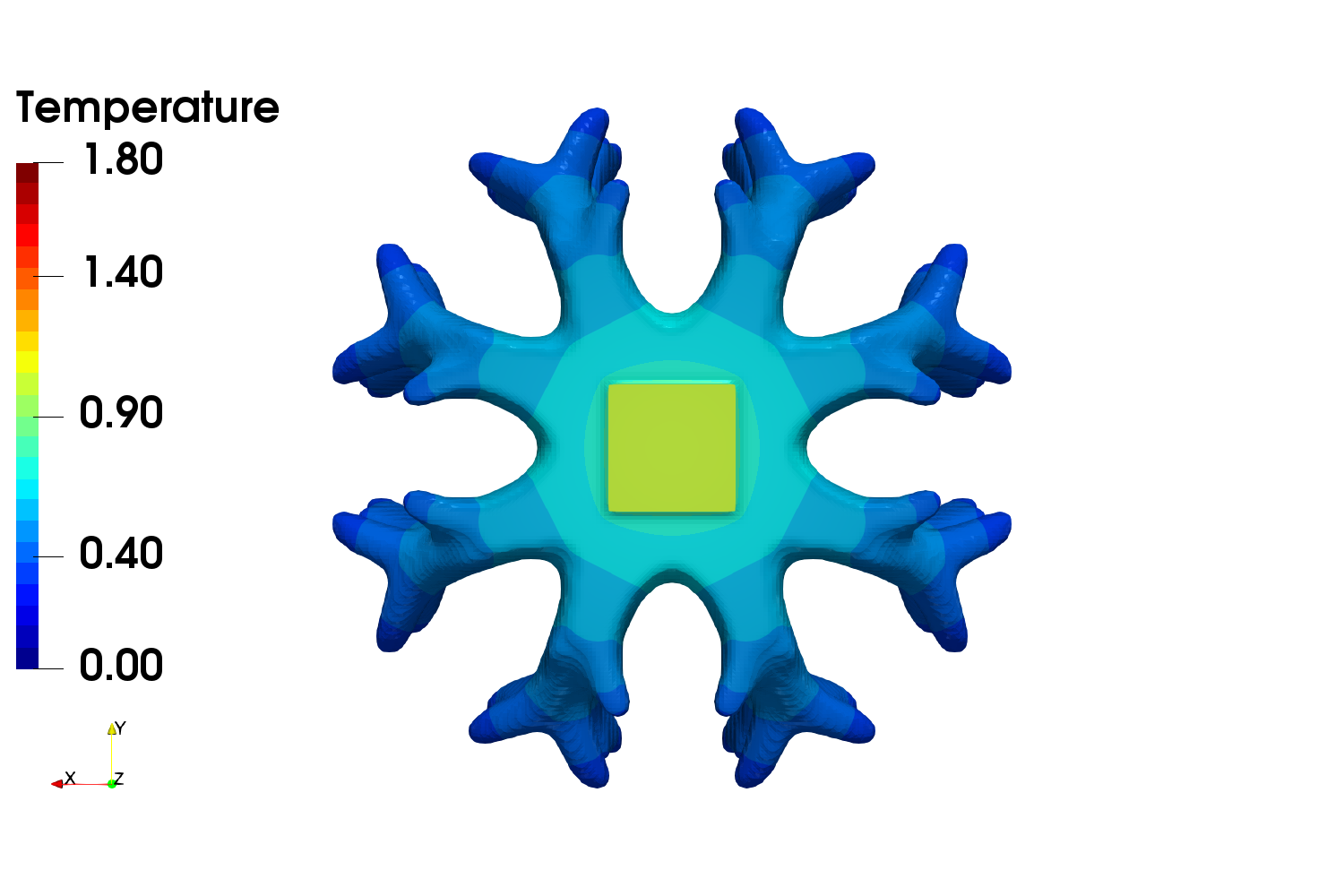}
        \caption{$\alpha=10^{6}K^{-1}$}
        \label{fig:shulitiger2}
    \end{subfigure}
    \caption{Bottom view - Temperature $[K]$ distribution in the optimized designs for $\alpha=\{10^{3},10^{6}\}$ $K^{-1}$. Results obtained with a mesh resolution of $160\times 160 \times 320$ elements, and performing $200$ optimization iterations considering a Navier-Stokes fluid flow followed by $300$ optimization iterations considering a Darcy fluid flow.}\label{fig:NS+D2}
\end{figure*}

\begin{table}[hbt]\centering\small
\caption{Computational time $[hours]$ required by the optimization analyses and final values of the objective function (i.e. $f$) for different $\alpha$ values $[K^{-1}]$. The results refer to the optimized designs shown in Fig. \ref{fig:NS+D1} and Fig. \ref{fig:NS+D2} obtained considering the hybrid approach described in Sec. \ref{sec:hybridnatconv}.}
\label{tab:topoptperformancehybrid}       
\begin{tabular}{llccc}
\hline\noalign{\smallskip}
 &  & \multicolumn{3}{c}{Time } \\
\cline{3-5}\noalign{\smallskip}
  &   & Navier-Stokes & Darcy & Total \\
$\alpha$  & $f$   & ($200$ iter) & ($300$ iter) & ($500$ iter) \\
\hline\noalign{\smallskip}
$10^{3}$ & $8.76$  & $20.0$ & $1.5$ & $21.5$ \\
$10^{4}$ & $8.17$  & $20.0$ & $1.7$ & $21.7$ \\
$10^{5}$ & $6.60$ & $31.6$ & $2.0$ & $33.6$ \\
$10^{6}$ & $4.78$ & $32.0$ & $2.5$ & $34.5$ \\
\hline\noalign{\smallskip}
\end{tabular}
\end{table}
Also in this case, in the numerical examples we considered four different values of the coefficient of thermal expansion $\alpha$ as in Sec. \ref{sec:numexdiffalpha}. In the first stage of the problem governed by the Navier-Stokes fluid model, the parameters and the problem settings were defined as by \citet{Alexandersen2016}. 
In the second stage, the parameters and the problem settings were defined as in Sec \ref{sec:numexdiffalpha}.
We considered one unified continuation scheme for the following parameters:
\begin{equation}\label{eq:contsch2}
\begin{split}
& q_{c} = \left\{\underbrace{0.881, 8.81,}_\mathrm{Navier-Stokes} \underbrace{88.1, 88.1, 881.0}_\mathrm{Darcy}\right\}\\
& q_{p} = \left\{\underbrace{8.0, 8.0,}_\mathrm{Navier-Stokes} \underbrace{8.0,  98.0, 998.0}_\mathrm{Darcy}\right\}\\
& \kappa_{s} = \left\{\underbrace{10^{5}, 10^{5},}_\mathrm{Navier-Stokes} \underbrace{10^{-5}, 10^{-6}, 10^{-7}}_\mathrm{Darcy}\right\}\, m^{2}
\end{split}
\end{equation}
where the parameters' values are updated every $100$ iterations.
It should be noted that in the first stage, $q_{c}$ interpolates the material conductivity as in the second stage, while $q_{p}$ interpolates the impermeability in the solid phase instead of the permeability between the solid and fluid phases as in the second stage.
Consequently, $\kappa_{s}$ in the first stage represents the impermeability of the solid material, while in the second stage it represents the permeability of the solid material.
More details regarding the interpolation schemes adopted in the first optimization stage where the Navier-Stokes fluid flow model is considered can be found in \cite{Alexandersen2016}.
Table \ref{tab:topoptperformancehybrid} lists the values of the objective function of the optimized designs, and the computational time required in the different design cases in each specific optimization stage.
By comparing the computational time required in the first and second stages, one appreciates the significant computational savings obtained by switching to a simplified fluid model in the second stage of the optimization analysis.
Fig. \ref{fig:NS+D1} and \ref{fig:NS+D2} show the optimized topologies and their associated temperature distributions in the fluid for $\alpha=\{10^{3},10^{6}\}$ $K^{-1}$. In particular, in the convection-dominated case (i.e. $\alpha=10^{6}$ $K^{-1}$) the final optimized topology strongly resembles the one presented by \citet{Alexandersen2016}. This is particularly evident if the optimized design is viewed from the bottom, as shown in Fig. \ref{fig:shulitiger2}. This result highlights the benefit of considering the hybrid approach proposed here, which consists in the ability of reducing significantly the computational effort required in the optimization process while achieving optimized designs that are very similar to those obtained considering only the Navier-Stokes fluid flow model.

\begin{table}\centering\small
\caption{Cross-check of the objective function for the verification of the designs of Fig. \ref{fig:NS+D1} using COMSOL. Text coloring between blue (cold) and red (hot) is scaled according to the designs with minimum and maximum performance for each analysis case (i.e. $\alpha$ value $[K^{-1}]$). Bold numbers highlight the minimum value of the objective function for a given analysis case.}
\label{tab:topoptperformancehybrid2}       
\begin{tabular}{lcccc}
\hline\noalign{\smallskip}
 & \multicolumn{4}{c}{Optimization $\alpha$}  \\
\cline{2-5}\noalign{\smallskip}
Analysis $\alpha$ & $10^{3}$ & $10^{4}$ & $10^{5}$ & $10^{6}$\\
\hline\noalign{\smallskip}
$10^{3}$  & $\textcolor{mycolor91}{5.62}$ & $\textcolor{mycolor92}{\bm{5.53}}$ & $\textcolor{mycolor93}{6.11}$ & $\textcolor{mycolor94}{6.89}$\\
$10^{4}$  & $\textcolor{mycolor101}{5.34}$ & $\textcolor{mycolor102}{\bm{5.15}}$ & $\textcolor{mycolor103}{5.53}$ & $\textcolor{mycolor104}{6.14}$  \\
$10^{5}$ & $\textcolor{mycolor111}{4.04}$ & $\textcolor{mycolor112}{3.96}$ & $\textcolor{mycolor113}{\bm{3.85}}$ & $\textcolor{mycolor114}{4.04}$ \\
$10^{6}$ & $\textcolor{mycolor121}{2.90}$ & $\textcolor{mycolor122}{2.96}$ & $\textcolor{mycolor123}{2.81}$ & $\textcolor{mycolor124}{\bm{2.76}}$ \\
\hline\noalign{\smallskip}
\end{tabular}
\end{table}
For verification, the optimized designs have again been analyzed using COMSOL Multiphysics 5.3 \citep{COMSOL}. As before, the designs have been analyzed for all the values of $\alpha$ considered in the optimization analyses, and the results of the cross-check are reported in Table \ref{tab:topoptperformancehybrid2}.
First, one observes that all the optimized designs, except the one obtained for $\alpha = 10^{3 }\, K^{-1}$, have a better performance than the designs discussed in Sec. \ref{sec:numexdiffalpha}, if analyzed with the same full Navier-Stokes flow model. Additionally, the designs obtained for $\alpha=\{10^{4},10^{5},10^{6}\}$ $K^{-1}$ perform at best under the same condition (i.e. same $\alpha$) for which they are designed for, even though the models considered for design and verification are different. The design obtained for $\alpha=10^{3}$ $K^{-1}$ shows a loss of performance if tested with the high-fidelity model, even though the difference in performance is only about $1.6\%$ with respect to the best performance observed in the verification phase for $\alpha=10^{3}$ $K^{-1}$.

\section{Application example - suspended heated cylinder} \label{sec:secondExample}

This example further explores the performance of the simplified model for more complex problems. More specifically, the problem treated is that of a heated cylinder suspended in a closed cavity. 

\subsection{Problem setup}

\begin{figure}\scriptsize
\centering
  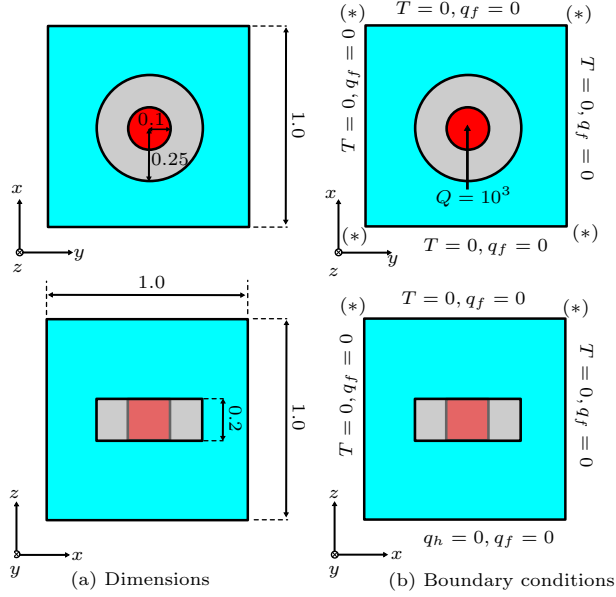\normalsize
\caption{Representation of the system considered in the second example, Section \ref{sec:secondExample}. The design domain is grey. The fluid domain is blue and the solid domain with volumetric heat source is red, neither of which are involved in the design. The pressure is zero at at the corners denoted by $(*)$.}
\label{fig:example2_setup}      
\end{figure}
Fig. \ref{fig:example2_setup} shows the problem setup, where the red cylinder represents the heat source, with radius of $0.1m$, height of $0.2m$ and a volumetric heat of $Q=10^{3}\,W/m^{3}$. The design domain is represented by the grey domain, which surrounds the heated cylinder with the same height and an outer radius $0.25m$.
The cylindrical setup is placed in the centre of a closed cavity with dimensions $1 \,m\times 1\,m\times 1\,m$. As for the benchmark example, the external vertical and top walls are kept at a constant temperature $T=0\, K$. The external wall at the bottom is insulated.
The boundary conditions and the geometry of the problem are both symmetric. Thus, in the computations we consider a quarter of the original domain with symmetry boundary conditions. The volume fraction is $15\%$, i.e. $g^{*}=0.15$ in Eq. \eqref{eq:constr}.
The conductivity of the solid material is $k_{s}=100\,W/mK$, while the conductivity of the fluid material is $k_{f}=1\,W/mK$. The remaining parameters are set as follows: $\rho_{0}=1\,kg/m^{3}$, $\mu=1\,Pa\,s$, $c_{p}=1\,J/kgK$. The problem is investigated for $\alpha = 10^6 K^{-1}$, the convective regime. The computational domain is a quarter of the presented, discretized using $160\times 160 \times 320$ elements, and the filter radius was set to 2.5 times the element size, i.e. $0.003125 \times 2.5 = 0.0078125 \, m$.

\subsection{Initial design}
For this problem, an initial design is considered. The initial design has four regularly-spaced radial fins with a total volume that equals the above-mentioned volume constraint. In order to set the fluid permeability, the analytical expression Eq. \eqref{eq:nusselt5} is used with $L=0.5m - 0.1m = 0.4m$ due to the extension of the inner heated cylinder from the centre. The temperature difference, $\Delta T$, is found using COMSOL simulations of the initial design. This is how we envision that a thermal engineer would estimate this in practice. The average surface temperature is $\Delta_T = 0.613 \,K$ yielding a final fluid permeability $\kappa_{f} = 0.000676 \,m^2$.

Since an initial design is supplied to the optimization algorithm, a different and shorter continuation strategy is applied:
\begin{equation}\label{eq:contsch3}
\begin{split}
& q_{c} = \left\{8.81, 88.1, 88.1 \right\}\\
& q_{p} = \left\{8.0,  98.0, 998.0\right\}\\
& \kappa_{s} = \left\{10^{-5}, 10^{-6}, 10^{-7}\right\}\,m^{2}
\end{split}
\end{equation}
where the parameters' values are updated every $100$ iterations. For the hybrid NS-D approach, Navier-Stokes is used for the first step with equivalent parameter values.

\subsection{Optimized designs}

\begin{figure*}
    \begin{subfigure}{0.32\textwidth}
    \centering
        \includegraphics[width=0.9\linewidth]{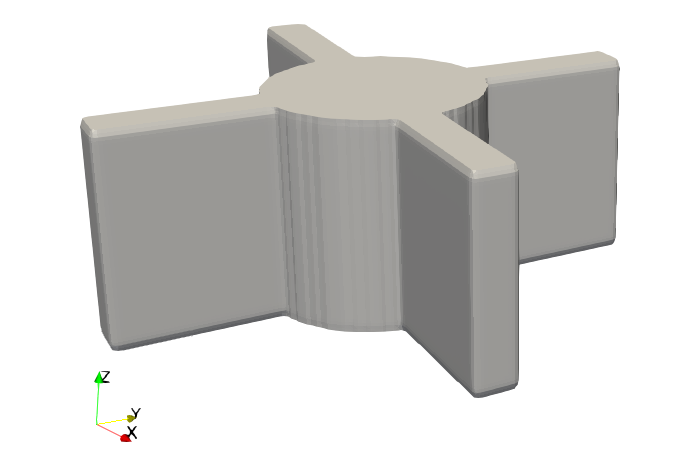}
        \\
        \includegraphics[width=0.9\linewidth]{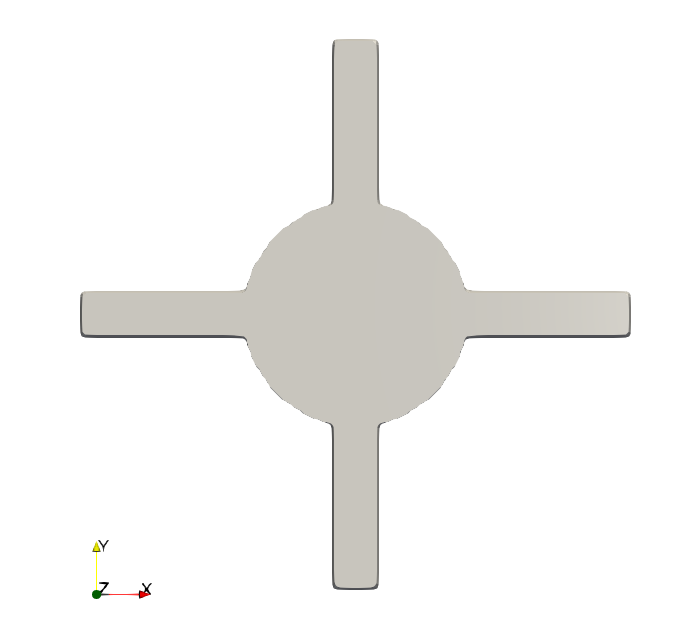}
        \caption{Initial}
        \label{fig:IC_designs_init}
    \end{subfigure}
    \begin{subfigure}{0.32\textwidth}
    \centering
        \includegraphics[width=0.9\linewidth]{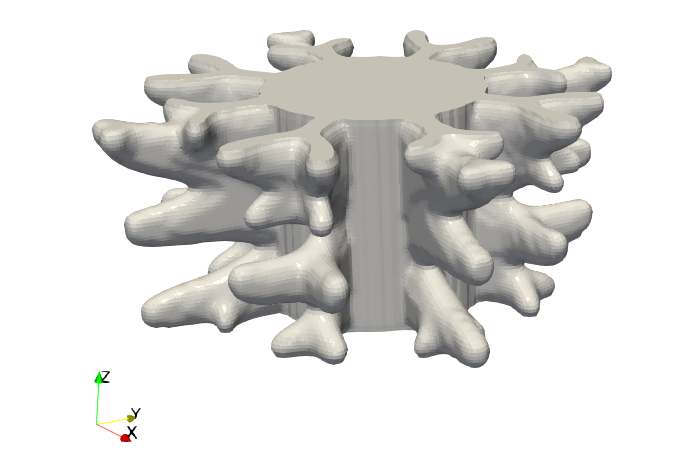}
        \\
        \includegraphics[width=0.9\linewidth]{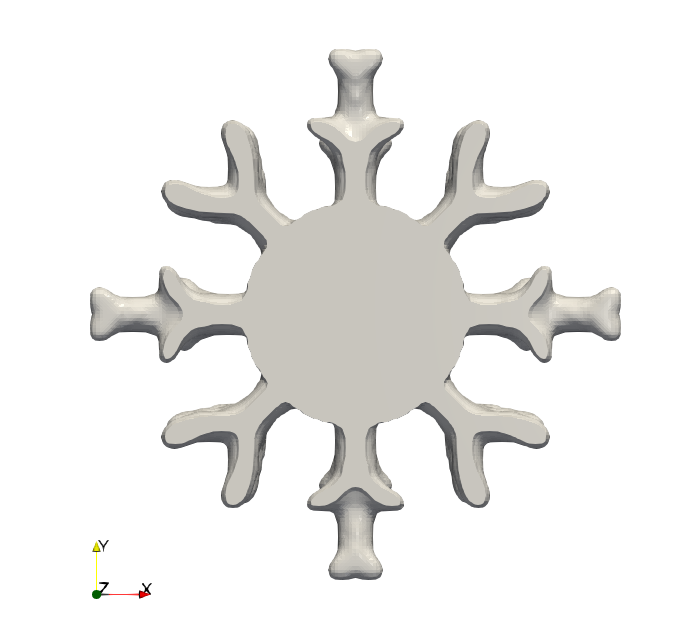}
        \caption{D}
        \label{fig:IC_designs_D}
    \end{subfigure}
    \begin{subfigure}{0.32\textwidth}
    \centering
        \includegraphics[width=0.9\linewidth]{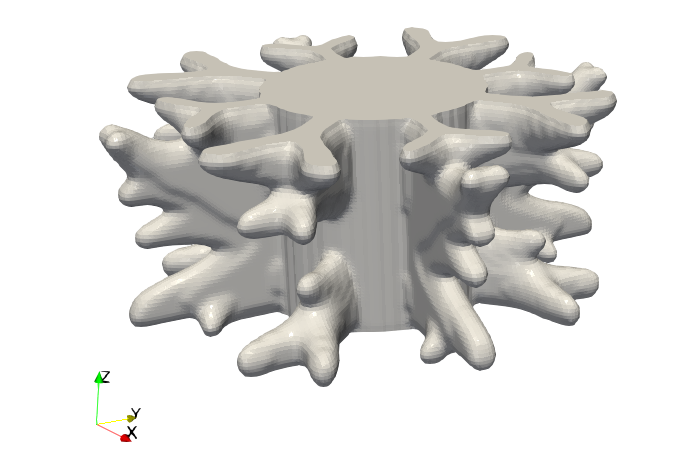}
        \\
        \includegraphics[width=0.9\linewidth]{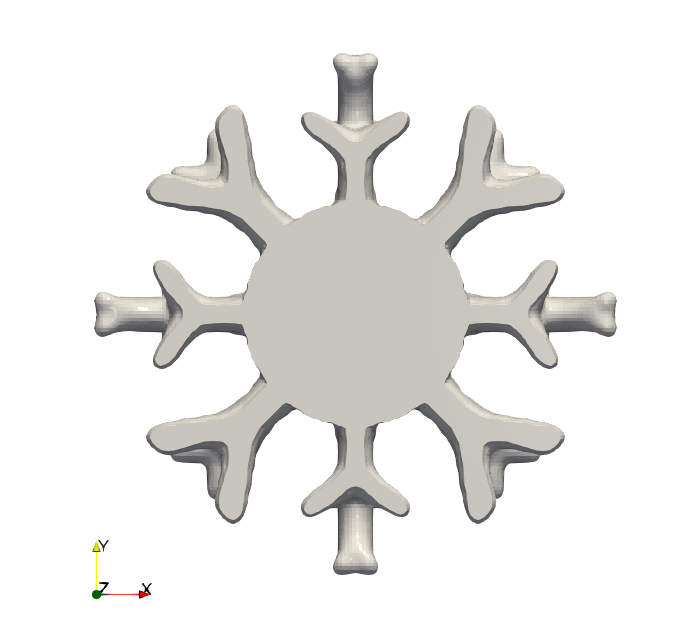}
        \caption{NS-D}
        \label{fig:IC_designs_NS-D}
    \end{subfigure}
    \caption{Initial and optimized designs for the second example. Results obtained with a mesh resolution of $160\times 160 \times 320$ elements, and performing $100$ optimization iterations per continuation step.}\label{fig:IC_designs}
\end{figure*}
Figure \ref{fig:IC_designs} shows the initial design, as well as optimised designs\footnote{In order to make the comparison to the reference design as fair as possible, the designs have been thresholded at a value of 0.5 in order to be volume preserving and ensure a similar volume of material as that of the reference.} using pure Darcy and hybrid approach with Navier-Stokes for 100 iterations and then Darcy for 200.
It can be seen that the optimised designs are quite different than the initial design. The optimised designs have eight radial members, where material has been carved from the original fins in order to build new features between them. The conducting members are seen to have elongated cross-sections and split into multiple secondary members.
The observed design features are perfectly in line with the previously presented results, both in the above benchmark example as well as for passive coolers for LED lamps \citep{Alexandersen2018}. However, the obtained characteristics are in direct contrast to those reported by \citet{Joo2018} for a very similar problem setup using a simplified NLC convection model.

\begin{figure}
	\centering
    \begin{subfigure}{0.5\textwidth}
    \centering
        \includegraphics[width=0.99\linewidth]{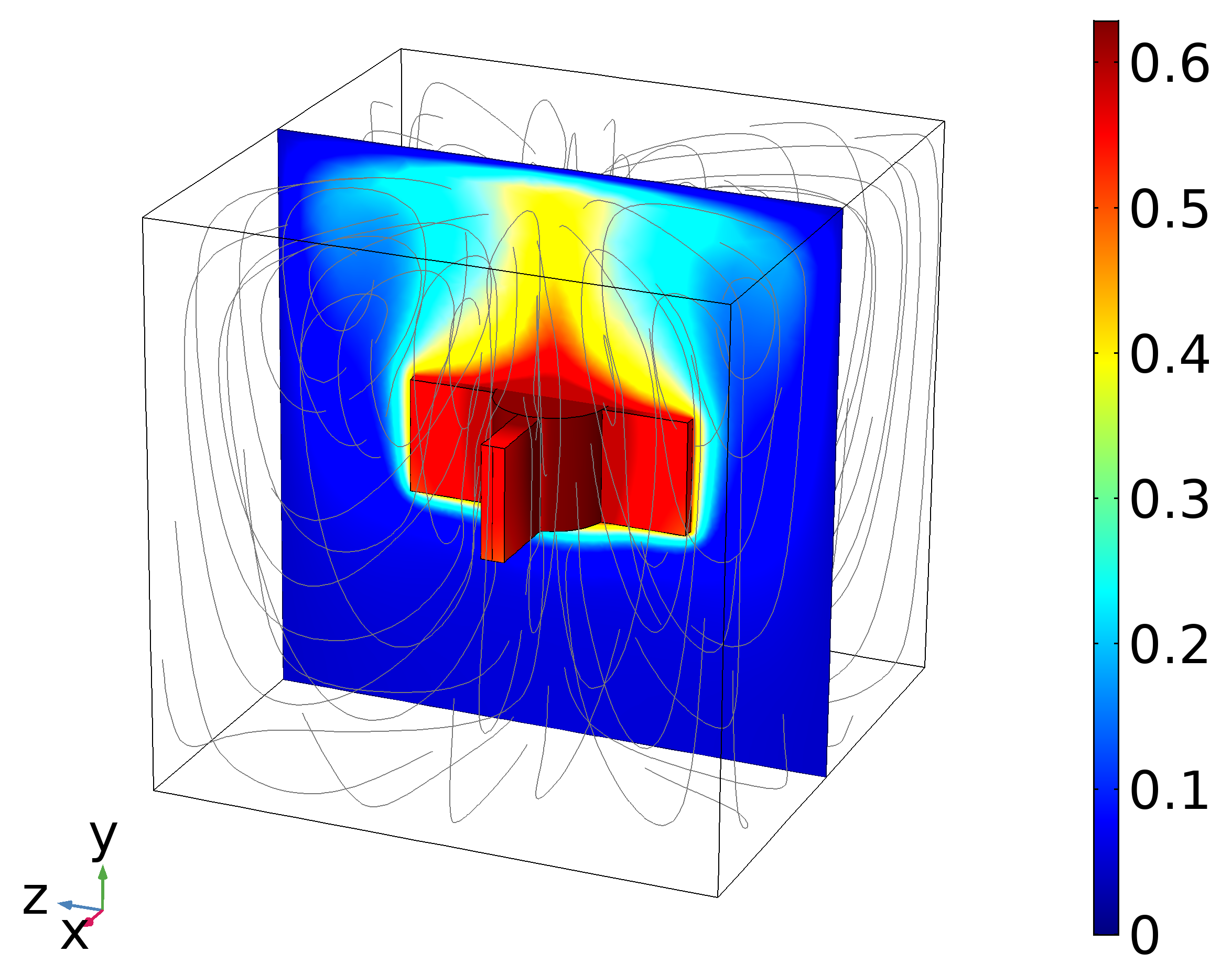}
        \caption{Reference}
        \label{fig:IC_COMSOL_init}
    \end{subfigure}
    \\
    \begin{subfigure}{0.5\textwidth}
    \centering
        \includegraphics[width=0.99\linewidth]{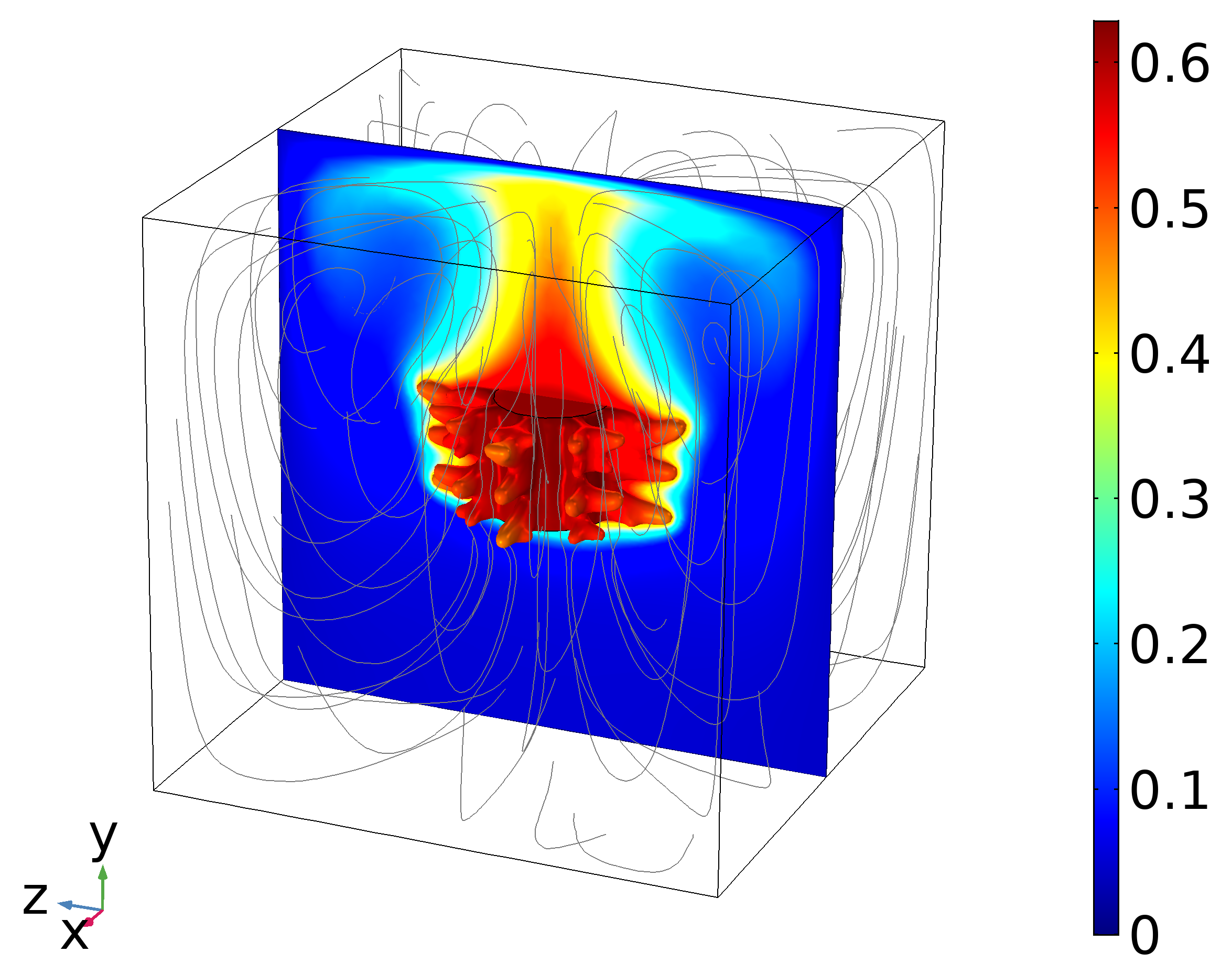}
        \caption{D}
        \label{fig:IC_COMSOL_D}
    \end{subfigure}
    \begin{subfigure}{0.5\textwidth}
    \centering
        \includegraphics[width=0.99\linewidth]{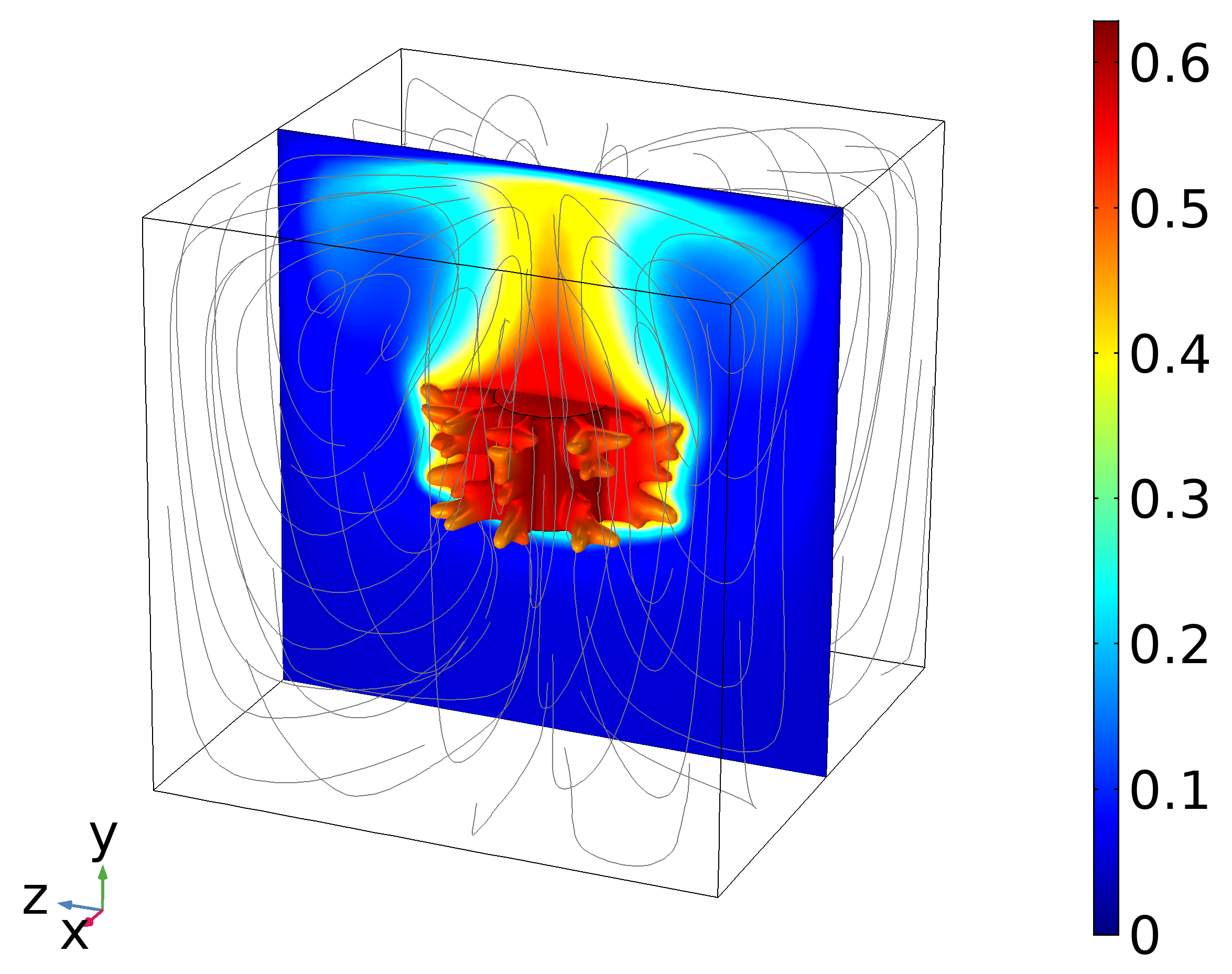}
        \caption{NS-D}
        \label{fig:IC_COMSOL_NS-D}
    \end{subfigure}
    \caption{Temperature fields and streamlines for the reference and optimized designs using COMSOL for the second example. The maximum of the scale is set to the maximum temperature of the best performing design (NS-D).}\label{fig:IC_COMSOL}
\end{figure}

The design obtained using pure Darcy flow is generally similar to that obtained using the hybrid NS-D approach. This indicates that good designs can generally be obtained using a well-tuned Darcy model. However, in order to verify the performance of the optimized designs, verification simulations using full Navier-Stokes model are performed using COMSOL. The models are discretized using $4-500,000$ elements with boundary layer refinement. The resulting temperature fields are shown in Figure \ref{fig:IC_COMSOL} and the performance is summarized in Table \ref{tab:IC_performance}.
Figure \ref{fig:IC_COMSOL} shows that the temperature fields and flow fields are generally very similar for the three designs. However, a slightly more compact plume appears above the optimized designs.
\begin{table}\centering\small
\caption{Performance of the reference and optimized designs for the second example. The average and maximum temperatures are calculated for the heated cylinder. The average temperature is proportional to the thermal compliance.}
\label{tab:IC_performance}       
\begin{tabular}{lccc}
\hline\noalign{\smallskip}
Design & Reference & D & NS-D  \\
\hline\noalign{\smallskip}
Volume $[m^3]$ & $0.00497$ & $0.00465$ & $0.00463$ \\
\hline\noalign{\smallskip}
$T_\text{avg}$ $[K]$ & $0.697$ & $0.625$ & $0.613$ \\
\hline\noalign{\smallskip}
$T_\text{max}$ $[K]$ & $0.714$ & $0.642$ & $0.629$ \\
\hline\noalign{\smallskip}
\end{tabular}
\end{table}
Table \ref{tab:IC_performance} shows that both optimized designs perform better than the reference design. The best performing design is obtained using the hybrid NS-D approach, having an average cylinder temperature that is $12\%$ lower than the reference design. The pure Darcy design is only slightly behind, having an average core temperature of only $2\%$ higher than the NS-D design.

\subsection{Computational time}
The computational time using 500 cores for the shown results was: 1 hour and 6 minutes for the pure Darcy optimization; and 8 hours and 39 minutes for the hybrid NS-D optimization.
Taking into account the relative performance of the Darcy design, this indicates that an improved design can be obtained in a very short time with a well-tuned Darcy model.
An estimate for the computational time using the full NS model is 20 hours, based on the time taken for the first 100 iterations of the hybrid NS-D approach. Thus, the pure Darcy optimization takes a mere $5.5\%$ of the time using the same number of cores.

The Darcy optimization was also run using 100 cores, taking 5 hours and 16 minutes, as well as using only 40 cores, taking 12 hours and 48 minutes.
Firstly, this shows the very good scalability of the framework, with $97\%$ and $93\%$ strong scaling efficiency for 100 and 500 cores, respectively, compared to 40.
Secondly, this shows that the proposed methodology is capable of providing an optimized design during the workday on a small computational cluster and overnight on a high-end dual-processor desktop.

\section{Conclusions}
\label{sec:end}

In this work we discuss a simplified approach for topology optimization of steady-state natural convection 3-D problems. The simplified approach discussed herein has been implemented in a PETSc-based framework for parallel computing that allows for optimizing high-resolution problems, significantly reducing the overall computational cost to $5-20\%$ in terms of core-hours compared to that required by the full Navier-Stokes model.
A significant contribution of the work discussed herein is the proposal of a simple, but effective, tuning procedure for the fictitious fluid permeability parameter derived from analytical expressions.

The proposed methodology has been applied to academic design cases, and the obtained designs show strong similarities to the results obtained by some of the authors in previous works where the full Navier-Stokes flow model was considered.
Also in this case, the optimized topologies are characterized by ``thermal trees'' that drive the heat away from the source towards cooler areas of the surrounding fluid. Moreover, the optimized topologies have the tendency to create longer branches in the conduction dominated case, and more contracted and branched topologies in the convection dominated case.

The optimized designs have been analyzed with the commercial finite element software COMSOL with a high-fidelity natural convection model.
The designs showed a modest loss of performance due to the discrepancy between the models adopted in the optimization and verification phases.
To improve the performance of the designs obtained, the simplified natural convection model has been deployed also in conjunction with a more accurate one, resulting in a hybrid optimization approach.
This led to final optimized designs with a better performance than those obtained considering the simplified fluid model only.  
The use of a hybrid approach also allowed for a significant reduction of the computational cost compared to the full Navier-Stokes model.  
The obtained designs strongly resemble those obtained considering the full model and could be used directly for further developments or as initial guesses for subsequent more sophisticated optimization analyses.

The suggested approach allows for a better characterization of the problem at hand compared to other even simpler natural convection models based on Newton's law of cooling.
The simplified model discussed herein overcomes some of the major limitations encountered with simpler approaches based on Newton's law of cooling, such as the over-prediction of the heat flux at the solid-fluid interface that leads to topologies characterized by very thin fluid channels and internal cavities.
However, due the assumptions used in deriving the simplified model, it also has limitations. Nevertheless, it appears that the model in general provides meaningful and well-performing designs - at least for the problems treated herein.

A question still remaining to be answered, is whether optimal heat sinks in natural convection are pin-, plate- or tree-like. Recent work on the simpler area-to-point conduction problem by \citet{Yan2018} proved that pin structures, and not tree structures, are optimal for that case.
The studies herein and by \citet{Alexandersen2016} seem to favor pin-like structures for conduction-dominated problems, which supports the findings of \citet{Yan2018}, and tree-like structures for convection-dominated problems.
However, to dig deeper into this question requires extensive parameter studies, which are out of the scope of the present work. Nevertheless, with the computational savings made possible with the suggested methodology, performing such extensive investigations has become significantly less demanding.

\section*{Acknowledgements}
The work presented in this paper was funded by the project ``Sapere Aude TOpTEN'' (Topology Optimization of Thermal ENergy systems) from the Danish Council for Independent Research (Grant No. DFF-4005-00320). The authors gratefully acknowledge this financial support.
The authors would like to thank the members of the TopOpt group for fruitful discussions.

\appendix
\section{Finite element formulation} \label{app:FEM}
The approximated problem takes the form:
\begin{equation}\label{eq:weakstable}
\begin{split}
&\text{Find ($P$,$T$) $\in$ $\mathcal{V}^{h}$ such that:}\\
& \int_{\Omega} \frac{\kappa}{\mu} \, \frac{\partial w}{\partial x_{i}}  \bigg( \frac{\partial P}{\partial x_{i}} + \rho_{0}\alpha g_{i} T \bigg) d\Omega
+ \int_{S_{u}} w q_{u}dS =0\\
&\int_{\Omega} w^{*} \left(-\rho_{0} c_{p} \frac{\kappa}{\mu} \bigg( \frac{\partial P}{\partial x_{i}} + \rho_{0}\alpha g_{i} T \bigg) \frac{\partial T}{\partial x_{i}} - Q \right)d\Omega +\\
& + \int_{\Omega} k\, \frac{\partial w}{\partial x_{i}}\, \frac{\partial T}{\partial x_{i}} d\Omega - \int_{S_{h}}  w\,q_{h}\,dS
=0,\quad \text{$\forall$ $w$ $\in$ $\mathcal{V}^{h}$}
\end{split}
\end{equation}
In Eq. \eqref{eq:weakstable}, a Petrov-Galerkin method has been used to stabilize the convective term of the weak form of the energy conservation equation. In particular, a streamline-upwind weight function has been adopted in the form:
\begin{equation}
w^{*} = w + \tau u_{i} \frac{\partial w}{\partial x_{i}}
\end{equation}
where $w$ is the weight function and $\tau$ the stabilization parameter. 
It should be noted that in Eq. \eqref{eq:weakstable} the second order derivative of the stabilization term has been neglected since we rely on linear finite elements.
In the work discussed herein, both the temperature and pressure fields have been discretized with the same shape functions. In particular for each element we have that:
\begin{equation}
\textbf{p}(\textbf{x}) = \textbf{N}(\textbf{x})^{T}\textbf{p}_{e}, \quad \textbf{t}(\textbf{x}) = \textbf{N}(\textbf{x})^{T}\textbf{t}_{e}
\end{equation}
where $\textbf{p}(\textbf{x})$ and $\textbf{t}(\textbf{x})$ are the interpolated pressure and temperature fields, $\textbf{p}_{e}$ and $\textbf{t}_{e}$ are the pressure and temperature nodal degrees of freedom, $\textbf{N}(\textbf{x})$ is the shape function vector and lastly $\textbf{x}$ is the coordinate vector.
Hence, the streamline-upwind weight function are defined as follows:
\begin{equation}
\textbf{N}^{*} = \textbf{N} + \tau \textbf{u}_{0}^{T} \textbf{B}
\end{equation}
As in \citep{Alexandersen20142D} and \citep{Alexandersen2016}, we adopt the stabilization parameter $\tau$ presented in \cite{Tezduyar2008a}:
\begin{equation}
\begin{split}
&\tau = \left( \frac{1}{\tau_{1}^{2}} + \frac{1}{\tau_{3}^{3}} \right)^{1/2}\\
& \tau_{1} = \dfrac{h_{e}}{ 2|| \textbf{u}_{0}||_{2}}=  \dfrac{h_{e}}{2}\left( \textbf{u}^{T}_{0}\textbf{u}_{0}\right)^{-\dfrac{1}{2}}\\
& \tau_{3} = \dfrac{h_{e}^{2}}{4}\\
\end{split}
\end{equation}
where the element length scale is $h_{e}=\sqrt{dx^{2}+dy^{2}+dz^{2}}$, with $dx$, $dy$, and $dz$ denoting the element dimensions in the three directions, and $\textbf{u}_{0}$ is the velocity vector evaluated in the element centroid.

After integrating Eq. \eqref{eq:weakstable}, we obtain the following system of equations:
\begin{equation}\label{eq:govequation}
\begin{bmatrix}
\textbf{P}_{p} & \textbf{P}_{t}\\
\textbf{0} & \textbf{T}^{1}_{t}-\textbf{T}^{2}_{t}-\textbf{T}^{3}_{t} 
\end{bmatrix}
\begin{bmatrix}
\textbf{p}\\
\textbf{t}
\end{bmatrix}=
\begin{bmatrix}
\textbf{f}_{p}\\
\textbf{f}_{t\,q_{h}}+\textbf{f}_{t\,Q}
\end{bmatrix}\\
\end{equation}
where the matrices and vectors of Eq. \eqref{eq:govequation} have been assembled from the element equivalents:
\begin{equation}
\begin{split}
& \textbf{P}_{p,e} = \int_{\Omega_{e}} \frac{\kappa}{\mu} \, \textbf{B}^{T}\textbf{B} \,d\Omega\\
&\textbf{P}_{t,e} = \int_{\Omega_{e}} \rho_{0}\alpha \textbf{B}^{T}\textbf{g}\textbf{N} \,d\Omega  \\
& \textbf{f}_{p,e} =\int_{\Gamma_{u}} \textbf{N}^{T} q_{f}d\Gamma
\end{split}
\end{equation}
and 
\begin{equation}
\begin{split}
& \textbf{T}^{1}_{t,e} = \int_{\Omega_{e}} k\, \textbf{B}^{T}\textbf{B} \,d\Omega\\
& \textbf{T}^{2}_{t,e} = \int_{\Omega_{e}} \rho_{0} c_{p} \frac{\kappa}{\mu}\rho_{0}\alpha \textbf{N}^{*T}\left(\left(\textbf{g}\textbf{N}\textbf{t}_{e}\right)^{T}\textbf{B}\right) \,d\Omega  \\ 
& \textbf{T}^{3}_{t,e} = \int_{\Omega_{e}} \rho_{0} c_{p} \frac{\kappa}{\mu}\textbf{N}^{*T}\left(\left(\textbf{B}\textbf{p}_{e}\right)^{T}\textbf{B} \right)\,d\Omega  \\
& \textbf{f}_{t\,q_{h},e} =\int_{\Gamma_{h}} \textbf{N}^{T} q_{h}d\Gamma\\
& \textbf{f}_{t\,Q,e} =\int_{\Omega_{e}} \textbf{N}^{*T} Q \,d\Omega\\
\end{split}
\end{equation}
The global system of equilibrium equations (Eq. \eqref{eq:govequation}) is posed in residual form as follows:
\begin{equation}\label{eq:finalsyseq}
\textbf{R}(\textbf{s})=\textbf{M}(\textbf{s})\,\textbf{s}-\textbf{f}
\end{equation}
where $\textbf{s}=[\textbf{p}^{T} \; \textbf{t}^{T}]^{T}$.
Eq. \eqref{eq:finalsyseq} represents a nonlinear system of equations. The details of the algorithm adopted to solve it will be given in Sec. \ref{sec:topopt}.



\end{document}